\documentclass[epj]{svjour}
\usepackage[dvipdf]{graphicx}
\usepackage{subfig}

\begin{document}

\title{Coexistence in the two-dimensional May--Leonard model with random rates}
\author{Qian He\inst{1} \and Mauro Mobilia\inst{2} \and 
        Uwe C. T\"auber\inst{1}}

\institute{Department of Physics, Virginia Tech,
     	Blacksburg, Virginia 24061-0435, U.S.A. \\ 
        \email{heq07@vt.edu; tauber@vt.edu} \and 
        Department of Applied Mathematics, School of Mathematics, 
        University of Leeds, Leeds LS2 9JT, U.K. \\
        \email{M.Mobilia@leeds.ac.uk}}

\date{Received: \today / Revised version: date}


\abstract{
We employ Monte Carlo simulations to numerically study the temporal evolution 
and transient oscillations of the population densities, the associated 
frequency power spectra, and the spatial correlation functions in the 
(quasi-)steady state in two-dimensional stochastic May--Leonard models
of mobile individuals, allowing for particle exchanges with nearest-neighbors 
and hopping onto empty sites. 
We therefore consider a class of four-state three-species cyclic predator-prey 
models whose total particle number is not conserved.
We demonstrate that quenched disorder in either the reaction or in the mobility
rates hardly impacts the dynamical evolution, the emergence and structure of
spiral patterns, or the mean extinction time in this system. 
We also show that direct particle pair exchange processes promote the formation
of regular spiral structures. 
Moreover, upon increasing the rates of mobility, we observe a remarkable change
in the extinction properties in the May--Leonard system (for small system 
sizes):
(1) As the mobility rate exceeds a threshold that separates a species 
coexistence (quasi-)steady state from an absorbing state, the mean extinction 
time as function of system size $N$ crosses over from a functional form 
$\sim e^{{\rm c}N} / N$ (where c is a constant) to a linear dependence; 
(2) the measured histogram of extinction times displays a corresponding 
crossover from an (approximately) exponential to a Gaussian distribution. 
The latter results are found to hold true also when the mobility rates are 
randomly distributed.
}

\maketitle

\section{Introduction}
\label{introd}

The rock--paper--scissors (RPS) system 
\cite{MaynardSmith,Hofbauer,Nowak,Szabo,MayLeonard} or, equivalently, 
three-species cyclic predator-prey models, have been widely studied in order to
understand biodiversity in ecology and biology 
\cite{May,Maynard,Michod,Sole,Neal}.
As a non-trivial model for cyclic competition \cite{Frachebourg}, RPS as well as
its variants constitute paradigmatic model systems to mathematically describe the
co-evolutionary dynamics of three coexisting species in cyclic competition, 
such as, e.g., realized in nature for three types of Californian lizards 
\cite{Sinervo,Zamudio}, and the coexistence of three strains of {\em E. coli} 
bacteria in microbial experiments \cite{Kerr}.
Among the various RPS variants, a particular model introduced by May and 
Leonard \cite{MayLeonard} has recently received much attention
\cite{Reichenbach1,Reichenbach2,Reichenbach4,Reichenbach3,Matti}, leading to 
novel results that have important implications for the formation and 
propagation of spatial patterns in ecological systems.

It has been demonstrated that in non-spatial model systems with RPS-type
competition, two of the three species typically evolve towards extinction in 
finite observable time \cite{Reichenbach3,Reichen,Berr}. 
However, when spatial degrees of freedom and species dispersal and associated 
interactions between nearest-neighbor particles are allowed, e.g., in 
lattice-based Monte Carlo simulations, the spatial fluctuations and 
correlations induce a coexistence state of all three species, and the emergence
of intriguing spatio-temporal structures
\cite{Frachebourg,Reichenbach1,Reichenbach2,Reichenbach4,Matti,BenNaim,Provata,Tainaka,Szabo2002,Perc,Tsekouras}.
In particular, the authors of 
Refs.~\cite{Reichenbach1,Reichenbach2,Reichenbach4} studied a stochastic 
two-dimensional version of the four-state RPS game (May--Leonard model) where 
the conservation law for the total population density is removed, and found 
that mobility (particle pair exchange together with hopping) has a critical
influence on species diversity. 
Significantly, when the mobility exceeds a well-defined threshold, the typical 
size of spiral patterns outgrows the system size, and eventually species
coexistence in the system is destroyed \cite{Reichenbach1,Matti}. 

Coexistence and competition of biological species are often crucially affected 
by environmental influences which include limited and randomly distributed 
natural resources, the availability of shelter, varying climate conditions, 
etc.
Therefore, it is important to understand the precise role of spatial 
inhomogeneity on the formation and development of biodiversity. 
In previous work concerned with a four-state RPS system with conserved total
particle number \cite{Qian}, we have found that spatially varying reaction 
rates have little effect on the dynamical evolution. 
In this paper, our goal is to numerically study the effect of spatial disorder 
on species coexistence in two-dimensional stochastic May--Leonard model 
variants. 
To this end, we implement the reaction (and mobility) rates for the 
May--Leonard dynamics as quenched random variables that for each lattice site
are independently drawn from a truncated Gaussian distribution, and 
subsequently held fixed during the simulation run.
We here explore (i) the self-organization of the population in the coexistence 
phase, (ii) compute the spatio-temporal correlation functions, and (iii) 
investigate the statistics of species extinction times (for small system sizes).  
Our main results can be summarized as follows:

(1) We demonstrate that quenched spatial disorder has only minor effect on 
species coexistence in the May-Leonard model, which together with the results 
reported in Ref.~\cite{Qian}, shows that RPS models (in the presence or absence
of total particle number conservation) form a class of systems that are robust 
against environmental variability.
Remarkably, this statement is true even when spatial disorder affects the
particles' mobility which is known to drastically impact species coexistence.

(2) We study the combined effect of pair exchange and hopping processes, and 
demonstrate that the former are more important for the formation of robust 
spiral wave structures.

(3) We compute the extinction times, defined as the time when the first one of 
three species dies out, in (small) spatially extended systems.
We thus find that the mean extinction time (MET) increases sharply with system 
size $N$ when the mobility rate is low and the system is in the (long-lived 
metastable) coexistence state. 
However, once the mobility exceeds the threshold beyond which species 
extinction is prevalent, the MET function switches to a linear dependence on 
$N$. 
Correspondingly, the extinction time distribution is found to cross over from 
approximately exponential with prominent tail at large times, to a bell-shaped
near-Gaussian function. 

This paper is structured as follows: 
In Sec.~\ref{meanft}, we define the stochastic spatial May--Leonard model as a 
four-state spatial rock--paper--scissors (RPS) game without conservation law 
for the total particle number, and briefly discuss the well-established results
from the mean-field rate approximation approach.
In Sec.~\ref{modint}, we first explain how this stochastic spatial system is 
implemented in our Monte Carlo simulations on a two-dimensional lattice, and 
introduce the quantities of interest.
Then we present and analyze our simulation results.   
Finally, in Sec.~\ref{conclu} we conclude with a discussion and interpretation 
of our findings.

\section{Model and rate equations}
\label{meanft}

In the mean-field approximation, our spatial system reduces to the original 
May--Leonard model\footnote{At the mean-field level, the model considered here
corresponds to the original May--Leonard model ~\cite{MayLeonard} with
parameters $\alpha = 1$ and $\beta = 1 + \sigma / \mu$, and time measured in 
units of $1/\mu$.} \cite{MayLeonard}.
We let all populations live on a square lattice, with each lattice site 
occupied with at most a single individual.  
We therefore allow four states per site: three interacting particle species 
that we label $A$, $B$, and $C$, and an empty state $\emptyset$.
The model is defined through the following set of binary predation and 
offspring production reactions between the three particle species
\cite{MayLeonard,Reichenbach1,Reichenbach2}:
\begin{eqnarray}
  A + B &\to \emptyset + A \quad &{\rm with \ rate} \ \sigma \ ;  \nonumber \\
  B + C &\to \emptyset + B \quad &{\rm with \ rate} \ \sigma \ ; \label{react1}
  \\
  C + A &\to \emptyset + C \quad &{\rm with \ rate} \ \sigma \ ; \nonumber \\
  X + \emptyset &\to X + X \quad &{\rm with \ rate} \ \mu \ , \label{react2}
\end{eqnarray}
where $X\in(A,B,C)$ refers to any one of the three species. 
Note that in contrast with the conventional rock--paper--scissors model 
\cite{Reichenbach3,Reichen,Qian,GSzabo,Szolnoki,Claussen}, the total particle 
number is {\em not} conserved by these reactions, owing to the separation of 
predation and reproduction processes.
In addition, in our spatially-extended system we consider the nearest-neighbor
particle exchange and hopping processes on a two-dimensional square lattice 
(with periodic boundary conditions; here again $X, Y = A, B, C$):
\begin{eqnarray}
  X + Y &\to Y + X  \quad &{\rm exchange \ rate} \ \epsilon \ ; \label{exch} \\
  X + \emptyset &\to \emptyset + X \quad &{\rm hopping \ rate} \ D \ . 
  \label{hopp}
\end{eqnarray}
It is worth mentioning that the models studied in 
Refs.~\cite{Reichenbach1,Reichenbach2,Reichenbach4} do not separate the hopping
process (\ref{hopp}) from pair exchange (\ref{exch}); i.e., $\epsilon = D$. 
Therefore, when letting $\epsilon = D$ and in the absence of any quenched 
disorder, our spatial model coincides with the one investigated in 
Refs.~\cite{Reichenbach1,Reichenbach2,Reichenbach4}.

May and Leonard \cite{MayLeonard} studied the associated deterministic 
mean-field rate equations and obtained the temporal evolution of the population
densities. 
Let $a(t)$, $b(t)$, and $c(t)$ represent the population densities 
(concentrations) of species $A$, $B$, and $C$, respectively.  
Since at most one individual is allowed on each site in the simulation, within 
the mean-field approximation, the overall population density $\rho(t) = 
a(t) + b(t) + c(t)$ restricts the reproduction processes (\ref{react2}).
Therefore, the corresponding rate equations are
\begin{eqnarray}
\label{RE}
  \partial_t \, a(t) &=& a(t) 
  \left[ \mu \, (1 - \rho(t)) - \sigma \, c(t) \right] \ , \nonumber \\
  \partial_t \, b(t) &=& b(t) 
  \left[ \mu \, (1 - \rho(t)) - \sigma \, a(t) \right] \ , \label{rateq} \\
  \partial_t \, c(t) &=& c(t) 
  \left[ \mu \, (1 - \rho(t)) - \sigma \, b(t) \right] \ . \nonumber
\end{eqnarray}

The coupled rate equations (\ref{rateq}) yield four linearly unstable absorbing
states $(a,b,c) = (0,0,0)$, $(1,0,0)$, $(0,1,0)$, and $(0,0,1)$, and one
reactive fixed point $(a^*, b^*, c^*) = \frac{\rho^*}{3} (1,1,1)$,
where $\rho^* = \frac{3 \mu}{3 \mu + \sigma}$, representing coexistence 
between the three species. 
Linearizing around the coexistence fixed point leads to
\begin{equation}
  \left( \begin{array}{c} \partial_t \, {\delta a} \\ 
  \partial_t \, {\delta b} \\ \partial_t \, {\delta c} \\ \end{array} \right) 
  = L \left( \begin{array}{c} \delta a \\ \delta b \\ \delta c \\ \end{array} 
  \right) \ ,
\end{equation}
where $\delta a(t) = a(t)-a^*$, $\delta b(t) = b(t)-b^*$, and 
$\delta c(t) = c(t)-c^*$, and with the linear stability matrix $L$
\begin{equation}
\label{L}
  L = \frac{-\mu}{3 \mu + \sigma} \left( \begin{array}{ccc}  
  \mu & \mu & \mu + \sigma \\ \mu + \sigma & \mu & \mu \\ 
  \mu & \mu + \sigma & \mu \end{array} \right) \ .
\end{equation}
Its eigenvalues are $\lambda_1 = -\mu$, and $\lambda_{2,3} = 
\frac{\mu \sigma}{2 (3 \mu + \sigma)}[1 \pm \sqrt{3} i]$, which demonstrates 
that the fixed point is locally stable only in one direction of parameter space
(the eigenvector associated with the negative eigenvalue $\lambda_1$), and
generally linearly unstable. 
As elaborated in Refs.~\cite{Reichenbach1,Reichenbach2,Reichenbach4}, the 
system dynamics quickly approaches an invariant manifold associated with the 
rate equations (\ref{RE}).  
In the neighborhood of the unstable interior fixed point $(a^*,b^*,c^*)$, the 
invariant manifold is tangent to the plane normal to the eigenvector of $L$ 
associated with $\lambda_1$ \cite{Reichenbach4}. 
On this invariant manifold, the trajectories approach the absorbing boundaries
of the phase portrait where they linger and form a heteroclinic cycle 
\cite{Hofbauer,MayLeonard}. 
In this case, any chance fluctuations can cause species extinction by deviating
the trajectories toward the absorbing boundaries.
From the imaginary part of the complex conjugate eigenvalues $\lambda_{2,3}$,
we infer the characteristic oscillation frequency 
$\omega = \sigma \rho^* / 2 \sqrt{3}$.

\section{Monte Carlo simulation results for the spatially-extended May-Leonard
         model}
\label{modint}

We investigate the two-dimensional May--Leonard model, i.e., the four--state 
stochastic RPS game defined by the reactions (\ref{react1},\ref{react2}) (which
do not conserve the total particle number) on a two-dimensional lattice 
(typically with $N = 256 \times 256$ sites) with periodic boundary conditions, 
subject to the nearest-neighbor exchange (\ref{exch}) and hopping (\ref{hopp}) 
processes. 
In order to mimic finite local carrying capacities, we impose a maximum 
occupancy number of one particle (of either species) per lattice site.  
When investigating the effect of spatial disorder on the model, we will treat 
one of the rates ($\mu, \sigma, \epsilon$, or $D$) at each lattice site as a 
random number drawn from a normalized Gaussian distribution truncated at one
standard deviation on both sides.
For example, $\mu \sim N(m,n)$ implies that the rate $\mu$ is picked from a 
truncated normal distribution on the interval $[m-n, m+n]$, centered at the 
value $m$ with standard deviation $n < m$.
In practice, a value of $\mu$ is drawn from $N(m,n)$ for each site on the 
lattice and attached to the corresponding site at the beginning of each single 
Monte Carlo run. 
The rate values remain unchanged for all sites until the next run is initiated.
Therefore, in our model, the randomized rates pertain to the lattice sites; but
are identical for any individual landing on a given site for each single run.

At each simulation step, an individual of any species on the lattice is 
selected randomly; then one of its four nearest-neighbor sites, which might be 
empty or occupied by one particle of either three species, is selected at 
random.
Subsequently, the particles undergo the pair reaction (\ref{react1}), 
reproduction (\ref{react2}), exchange (\ref{exch}), or hopping (\ref{hopp}) 
processes, according to the respective associated rates.  
Once on average each of the $P$ individual particles on the lattice has had a
chance to react, reproduce, exchange, or move, one Monte Carlo step (MCS) is 
completed; the infinitesimal simulation time step is thus  
$\delta t \sim P^{-1}$.

When studying the effect of quenched spatial disorder in the reaction and 
mobility rates, we shall focus on investigating a base model with (average) 
mobility rate set to $5$, since the corresponding pure system displays clearly
established spiral waves (see Fig.~\ref{fig1:snpsh2} below).
Whenever neither fixed rate values nor their distribution are specified below,
the following default rate values were implemented in the simulations:
$\mu = \sigma = 1$, and $\epsilon = D = 5$.
We shall characterize the emerging spatial structures through instantaneous
snapshots of the particle distribution in the lattice, and will depict the 
temporal evolution of population (spatially averaged) densities. 
Because of the underlying symmetry among the species $A$, $B$, and $C$, one 
representative population suffices and we here report results for the spatial 
average of local population number $n_A(j, t)$ of species $A$, i.e., the 
spatially averaged density 
$a(t) = \langle n_A(j, t) \rangle = \frac{1}{N} \sum_{j} n_A(j, t)$, where $j$ 
represents the site index.
We shall also obtain the associated Fourier transform 
\begin{equation}
  a(f) = \int a(t) \, e^{2 \pi i f t} \, dt \ ,
\label{fourtr}
\end{equation}
and compute the equal-time two-point correlation functions in the 
(quasi-)steady state,
\begin{equation}
  C_{AB}(x,t) = \langle n_A(j+x,t) \, n_B(j,t) \rangle - a(t) \, b(t) \ , 
\label{corfun}
\end{equation}
where $j$ denotes the site index, and similarly for $C_{AA}(x,t)$, etc.
For our typical system size of $N = 256 \times 256$ sites we never observed the
extinction state in our simulations. 
In fact, as discussed in Ref.~\cite{Reichenbach1} and below (see Sec.~3.3), in
this case one expects the extinction time to grow exponentially with the system
size.
Therefore, in order to access absorbing states and numerically measure the mean
extinction time (MET) $\bar{T}_{\rm ex} = \langle T_{\rm ex} \rangle$, where
$T_{\rm ex}$ is the extinction time for a single Monte Carlo run, and extract 
extinction time distributions, we must consider small systems of sizes $N = 25$
to $225$.

\subsection{Self-organization in the three-species coexistence phase}

\begin{figure*}[!t]
\label{Fig1}
\begin{center}
\subfloat[$\epsilon=0.1, D=0.1$]{\label{fig1:snpsh1}
\includegraphics[width=0.25\textwidth]{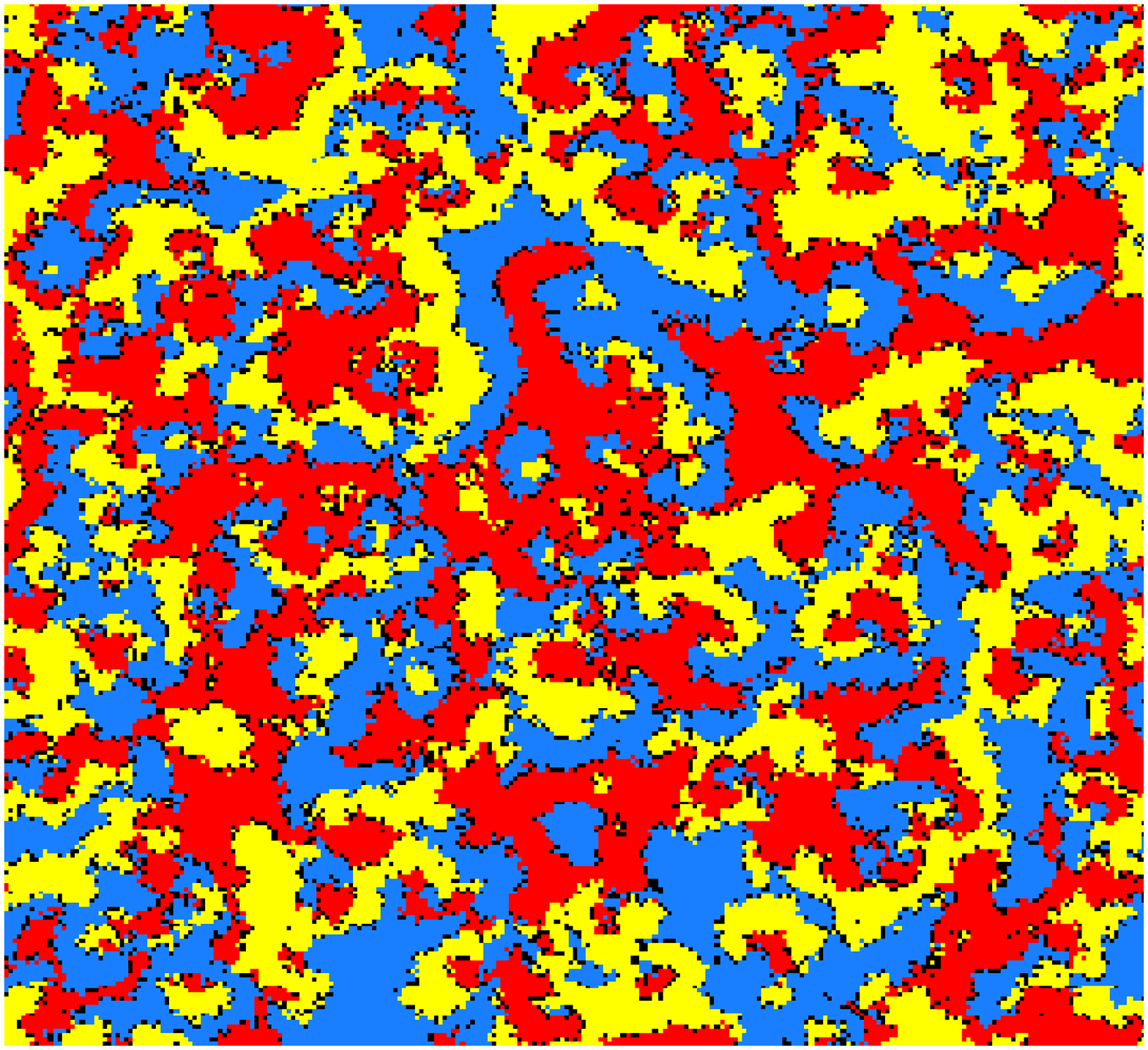}} 
\subfloat[$\epsilon=5, D=5$]{\label{fig1:snpsh2}
\includegraphics[width=0.25\textwidth]{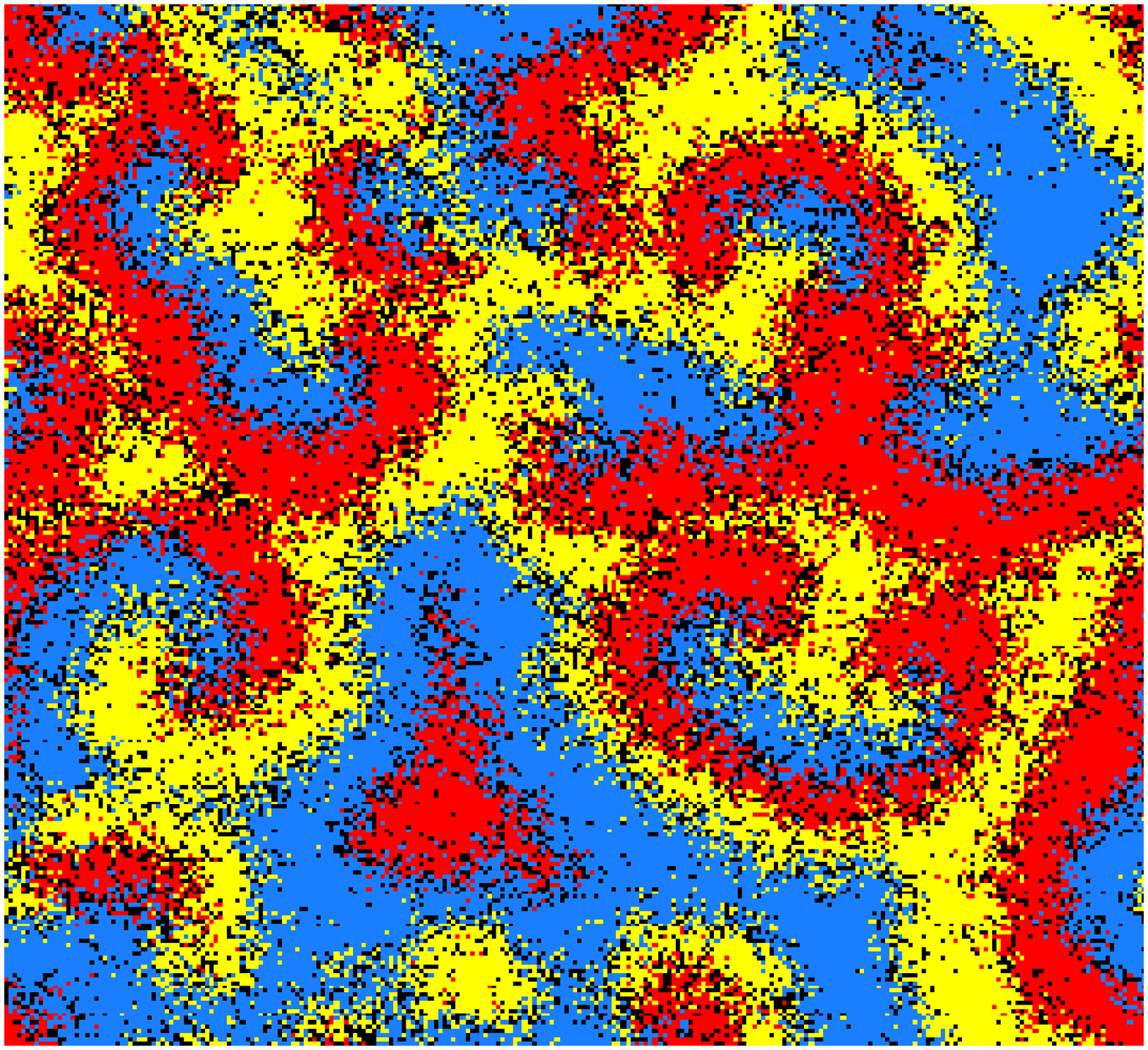}} 
\subfloat[$\epsilon=25, D=25$]{\label{fig1:snpsh3}
\includegraphics[width=0.25\textwidth]{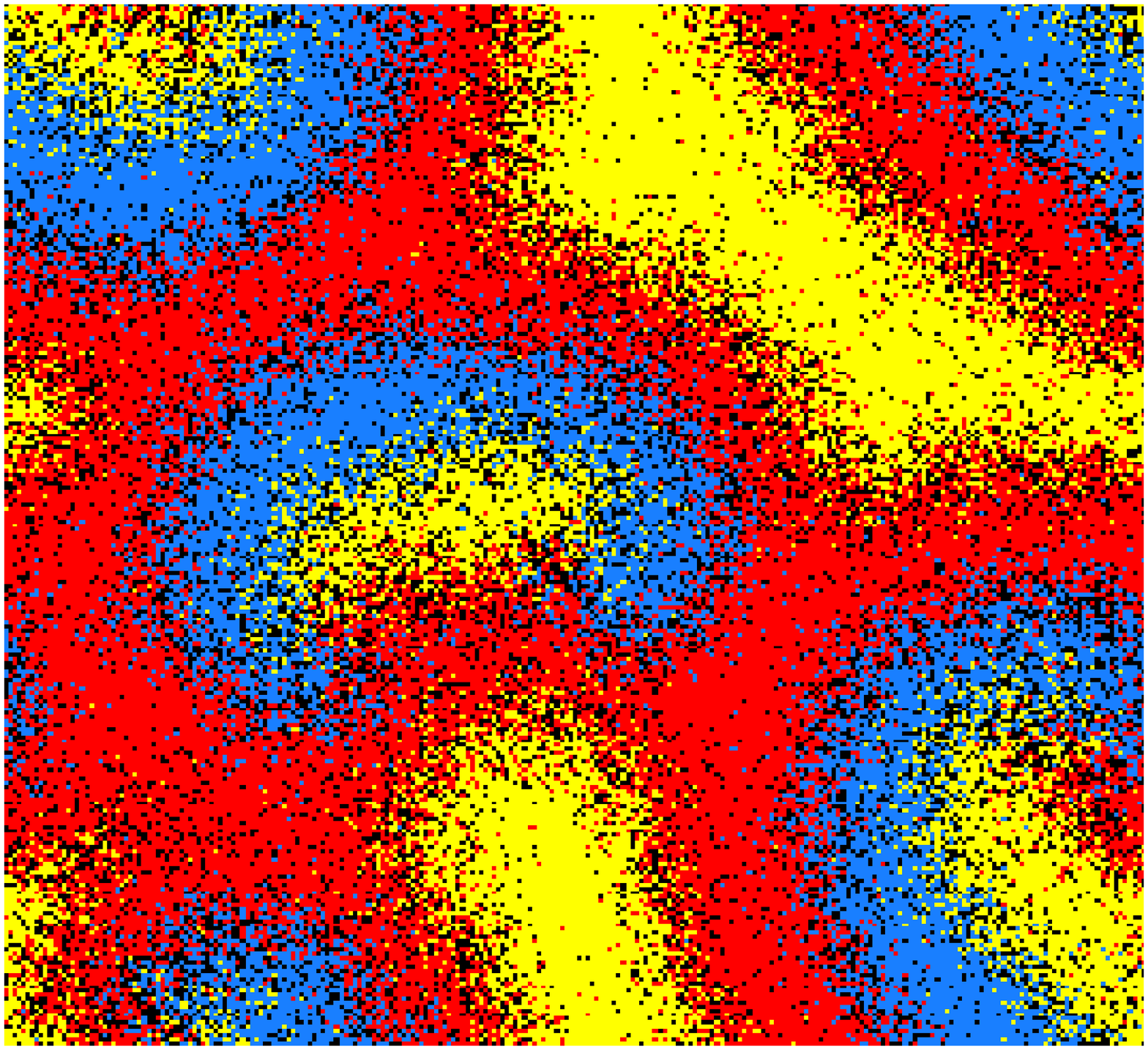}} \\
\subfloat[$\sigma \sim N(1, 0.5)$]{\label{fig1:snpsh4}
\includegraphics[width=0.25\textwidth]{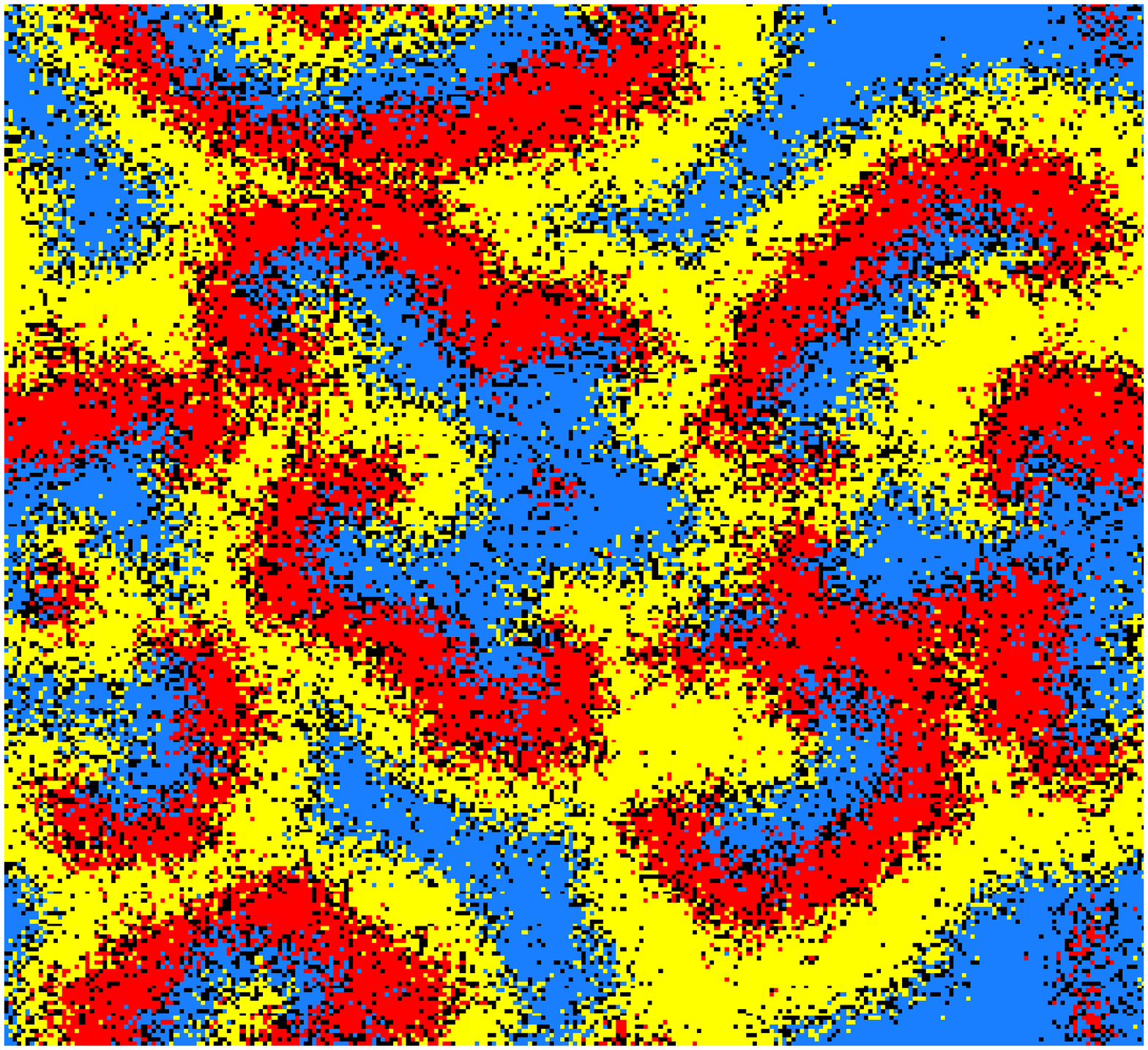}} 
\smallskip \subfloat[$\mu \sim N(1, 0.5)$]{\label{fig1:snpsh5}
\includegraphics[width=0.25\textwidth]{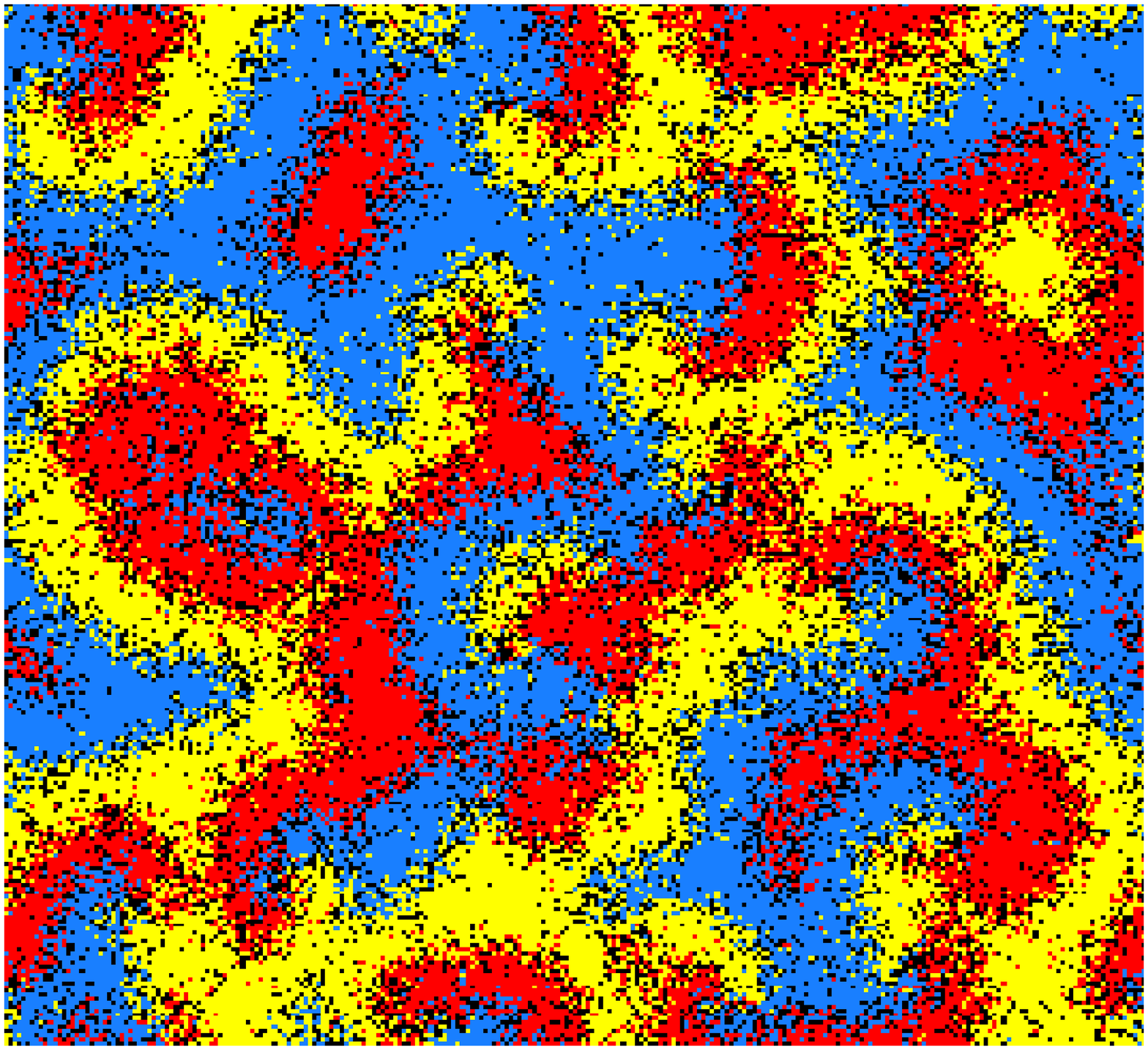}} 
\smallskip \subfloat[ $\epsilon, D\sim N(5, 4)$]{\label{fig1:snpsh6}
\includegraphics[width=0.25\textwidth]{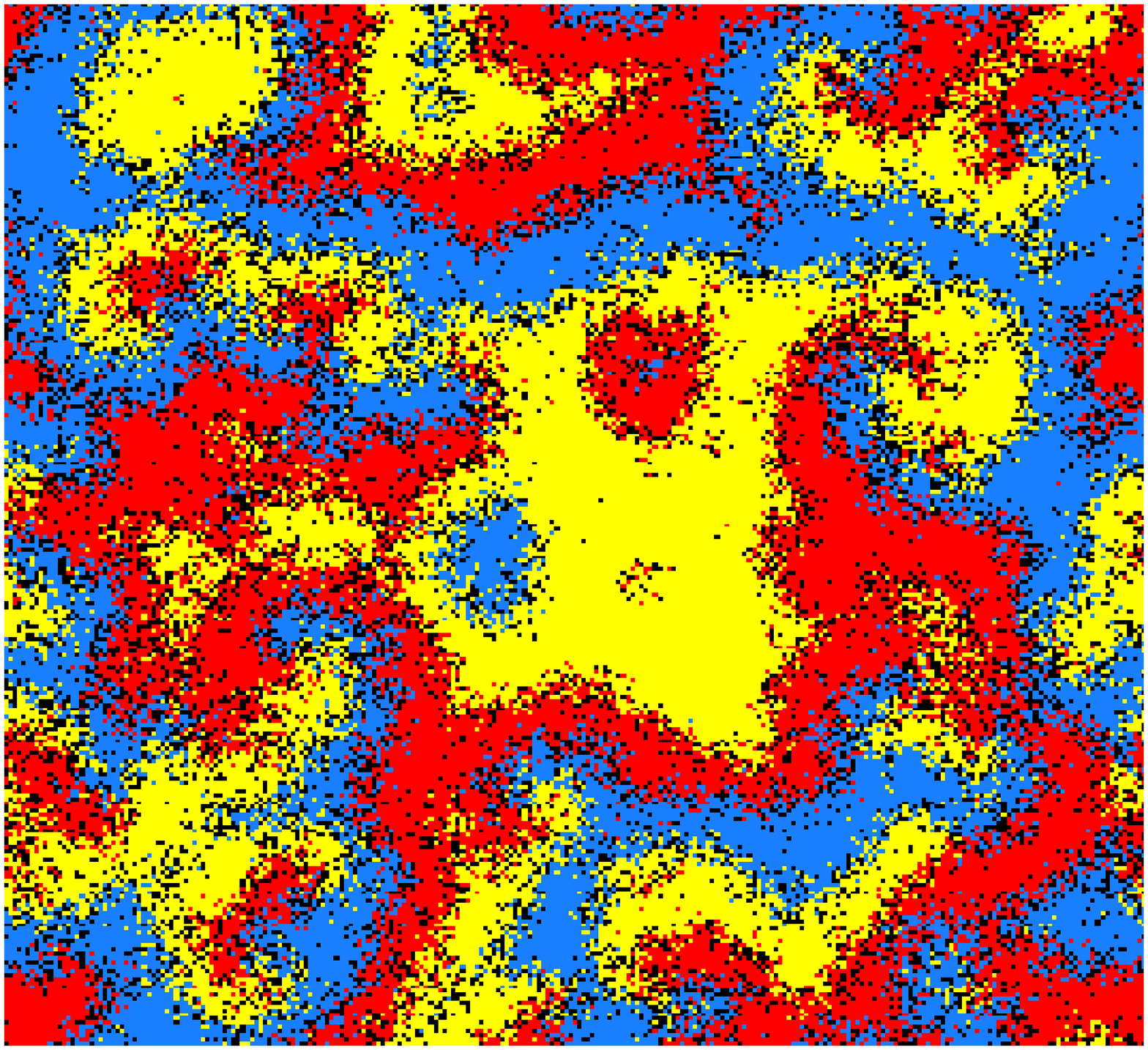}} \\
\subfloat[$\epsilon=0, D=5$]{\label{fig1:snpsh7}
\includegraphics[width=0.25\textwidth]{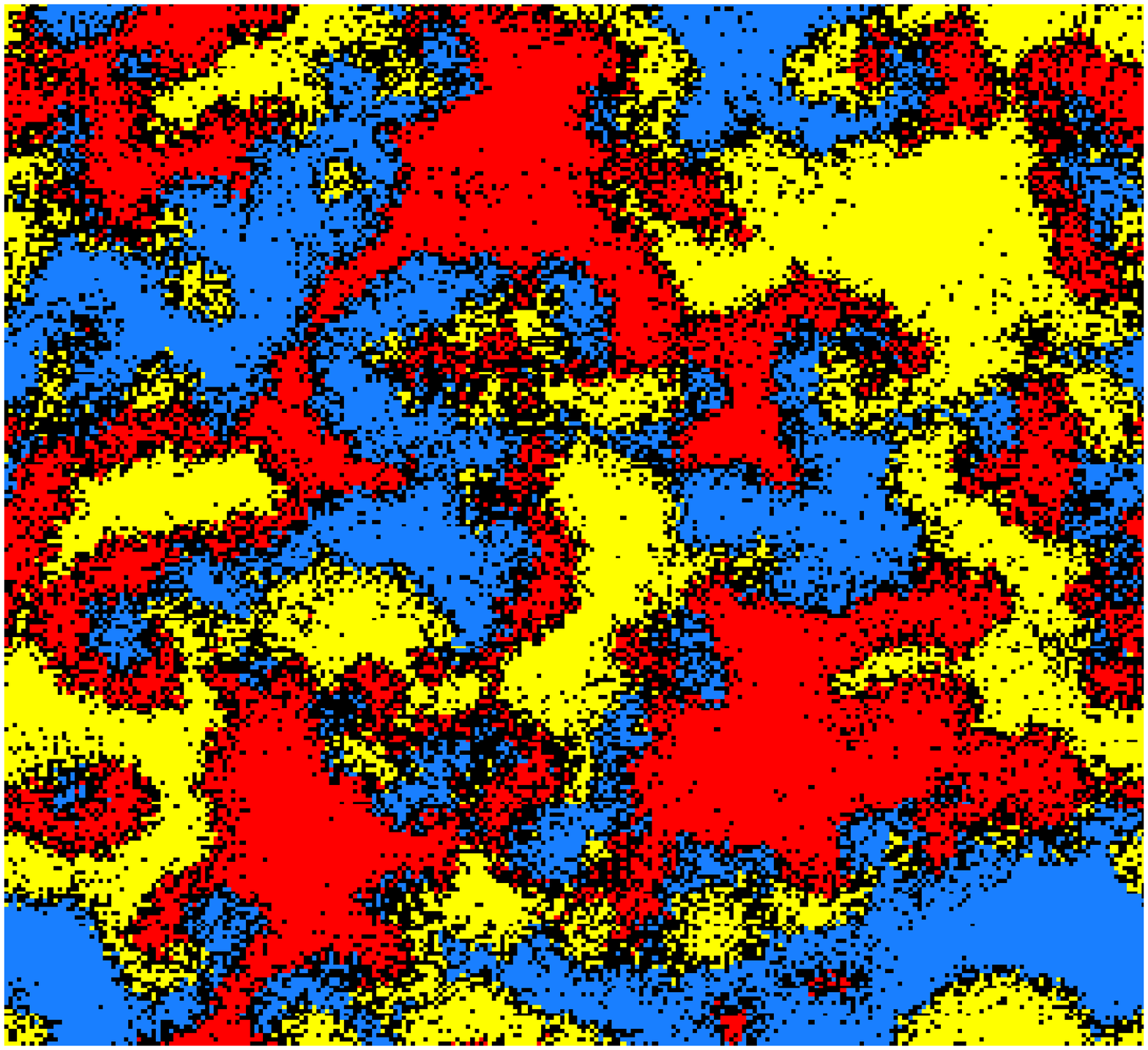}} 
\subfloat[ $\epsilon=5$, $D=0$]{\label{fig1:snpsh8}
\includegraphics[width=0.25\textwidth]{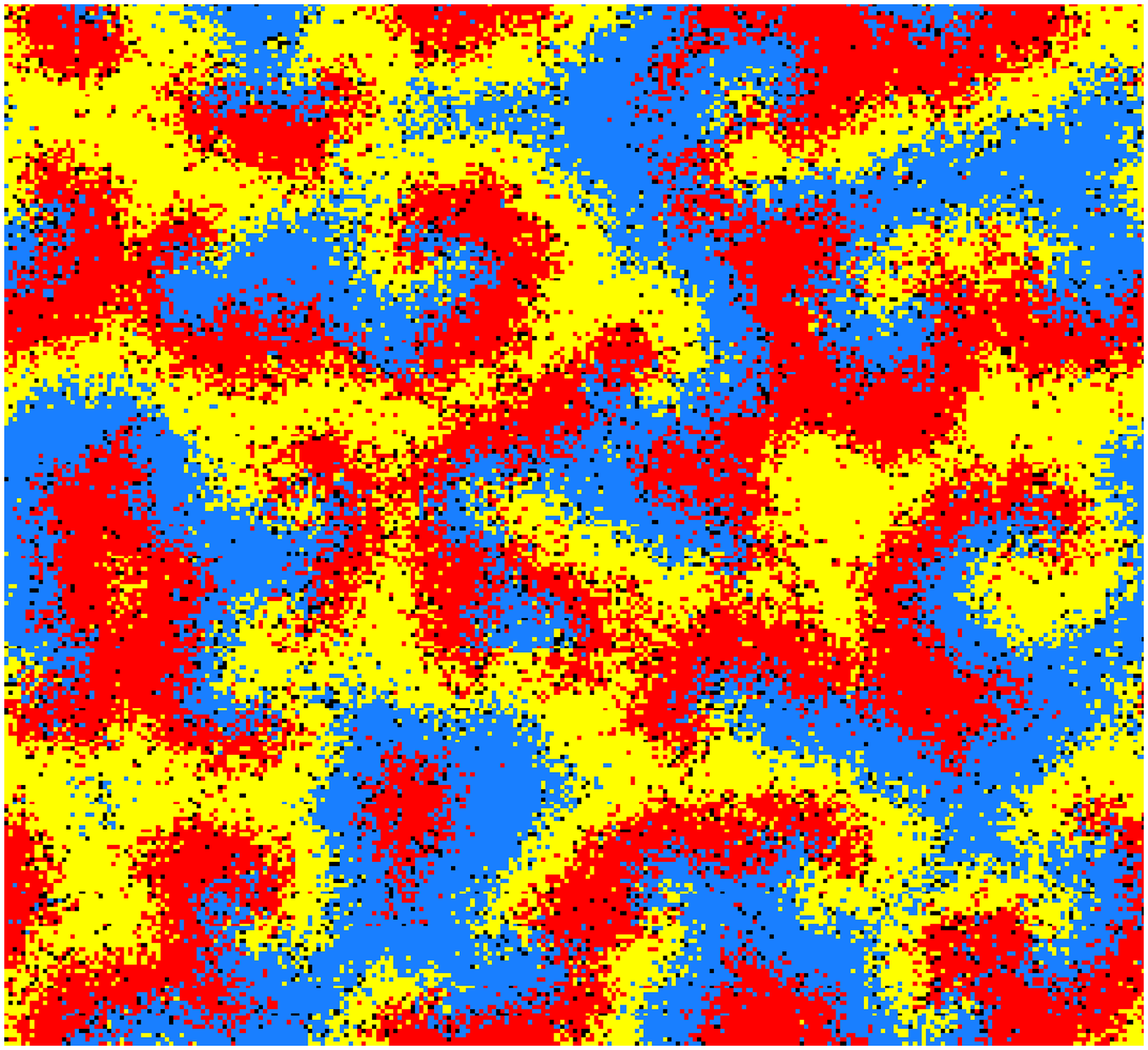}} 
\subfloat[ $\epsilon=0$, $D=25$]{\label{fig1:snpsh9}
\includegraphics[width=0.25\textwidth]{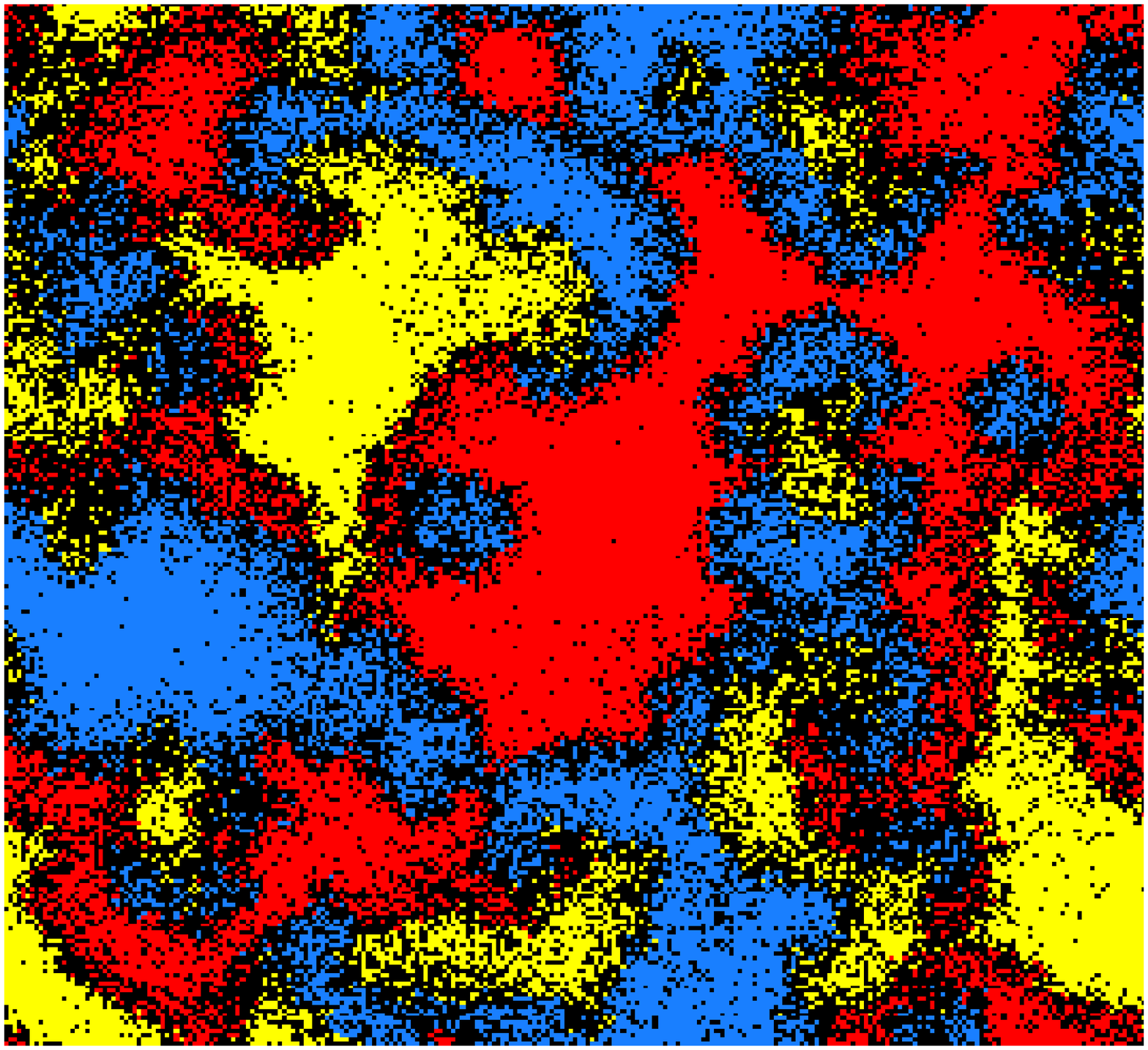}}
\end{center}
\caption{{\it (Color online.)} Snapshots of the spatial particle distribution 
  at $t = 1000$ MCS for a system with $N = 256 \times 256$ sites with equal 
  initial densities $a(0) = b(0) = c(0) = 1/3$.  
  If not specified otherwise, the corresponding default rate values are 
  implemented $\mu = \sigma = 1$, $\epsilon =  D = 5$.
  (red/gray: $A$, yellow/light gray: $B$, blue/dark gray: $C$, black: empty).}
\end{figure*}
In the first row of Fig.~1, we plot typical snapshots of the spatial particle 
distributions at $t = 1000$ MCS for various exchange and diffusion rates, as
indicated, while $\mu = \sigma = 1$ are held fixed.  
We observe in Fig.~\ref{fig1:snpsh1} that all three species coexist and a set 
of entangled spiral patterns forms in the system when the mobility rates are 
comparatively low. 
Upon increasing the mobility rates, the spiral patterns expand, see
Fig.~\ref{fig1:snpsh2}. 
When the spirals' typical size $\ell$ at last outgrows the lattice for large 
particle mobility (compare Fig.~\ref{fig1:snpsh1}--\ref{fig1:snpsh3}), we 
anticipate that the system essentially acquires the features of its 
zero-dimensional stochastic counterpart (see Sec.~3.3 and 
Ref.~ \cite{Reichenbach1}).
In that situation, the system evolves towards one of the three absorbing states
wherein two species become extinct, and the surviving species uniformly fills 
the lattice (uniform phase). 
In finite systems therefore, there exists a threshold set by the condition
$\ell \approx L = \sqrt{N}$ (in a square lattice) that separates the absorbing 
states from species coexistence.
Since the typical extent of the spirals grow diffusively as 
$\ell \sim \sqrt{2 \epsilon}, \sqrt{2 D}$, this transition should occur at some
critical value $M_c$ of the scaled effective mobility 
$M = 2 (\epsilon, D) / N$ \cite{Reichenbach1}.

In two-species predator-prey systems, quenched spatial disorder in the reaction
rates can markedly enhance both asymptotic particle densities in two-species 
predator-prey systems \cite{Ulrich}; yet this finding is not corroborated in 
three-species RPS models with conserved total population \cite{Qian}.
Thus we next explore the effect of spatial variability in the reaction as well
as in the mobility rates in the spatial stochastic May--Leonard model where the
conservation law for the total particle number has been removed, and where the 
mean-field dynamics is {\it not} characterized by neutrally stable orbits
\cite{Hofbauer,MayLeonard,Reichenbach3,Reichen}.
To this end, we introduce quenched spatial disorder by treating the rate on 
each site of the lattice as a random variable drawn from a truncated Gaussian
distribution.  
As shown in the snapshots in the second row of Fig.~1 (compare with 
Fig.~\ref{fig1:snpsh2}), the presence of spatial clustering can still be 
observed even though small noisy spiral structures dominate the system. 
Thus, the spatial disorder does not markedly affect the formation and 
occurence of spiral patterns. 

The third row of Fig.~1 shows snapshots of the system after removing either the
two-particle exchange process (\ref{exch}) or pure nearest-neighbor hopping
(\ref{hopp}). 
When exchange processes are not allowed, see Figs.~\ref{fig1:snpsh7} and 
\ref{fig1:snpsh9}, cluster formation is still observed, while the spiral waves 
become rather noisy, and it is worth noticing that one additional cluster type 
consisting of only empty sites appears in the system (discernible as black 
patches in Figs.~\ref{fig1:snpsh7} and \ref{fig1:snpsh9}), with measurable 
consequences on physical observables, as will be discussed below.
Moreover, in Fig.~\ref{fig1:snpsh8} where only exchange processes are allowed, 
the spiral pattern boundaries appear quite distinctly sharp, while the empty 
sites are randomly distributed rather than clustered.

From these results, we infer that the formation of the observed spiral patterns
is promoted by pair exchange processes. 
Furthermore, we have also verified that the above scenario is not affected when
the rates are randomly distributed and therefore remains robust against spatial
variability of the reaction rates.

\subsection{Time evolution and spatio-temporal correlation functions}

\begin{figure*}[!t]
\label{Fig2}
\begin{center}
\subfloat[Temporal evolution $a(t)$.]{\label{fig2:density}
\includegraphics[width=0.43\textwidth]{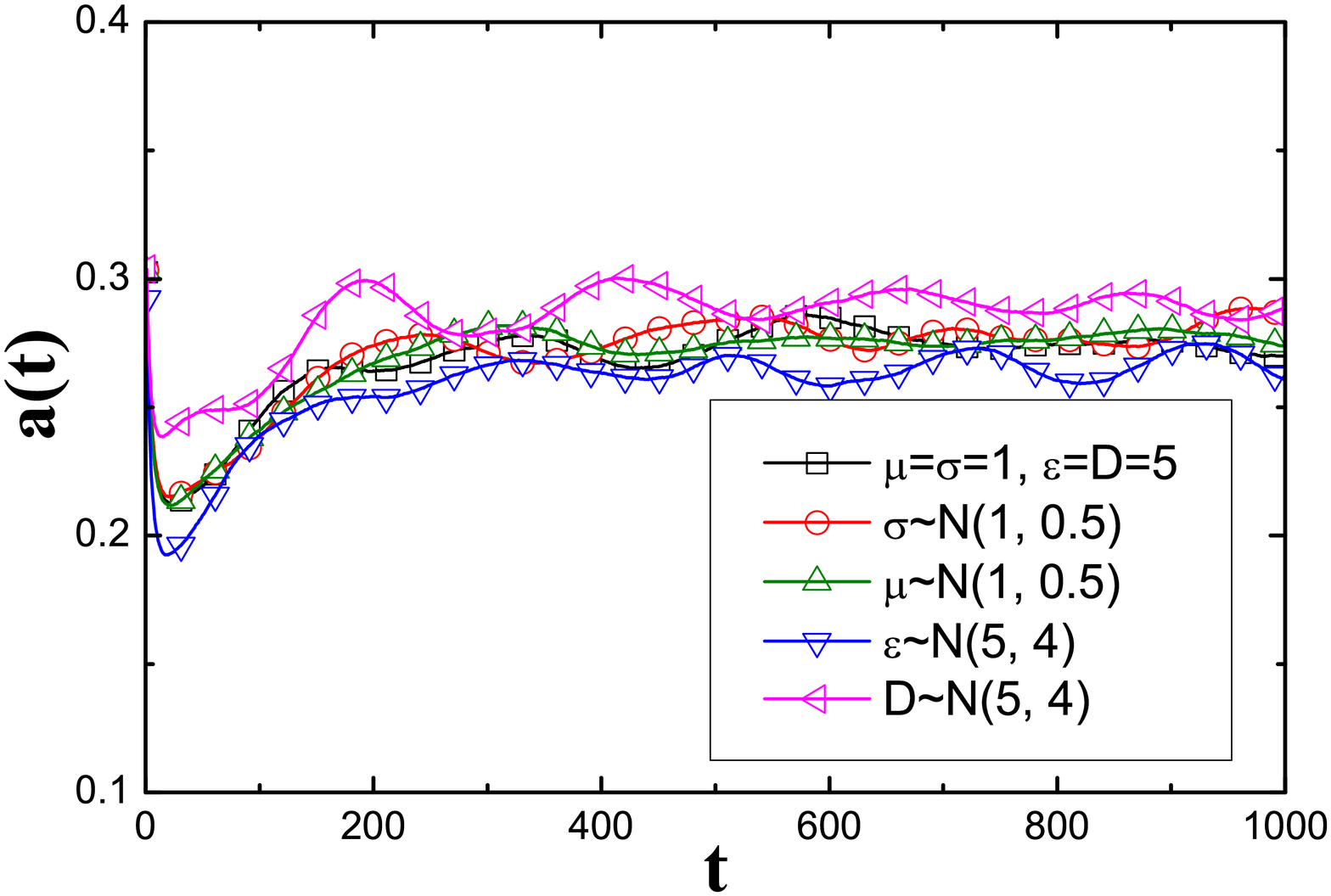}} 
\subfloat[Fourier transform $|a(f)|$.]{\label{fig2:FFT}
\includegraphics[width=0.43\textwidth]{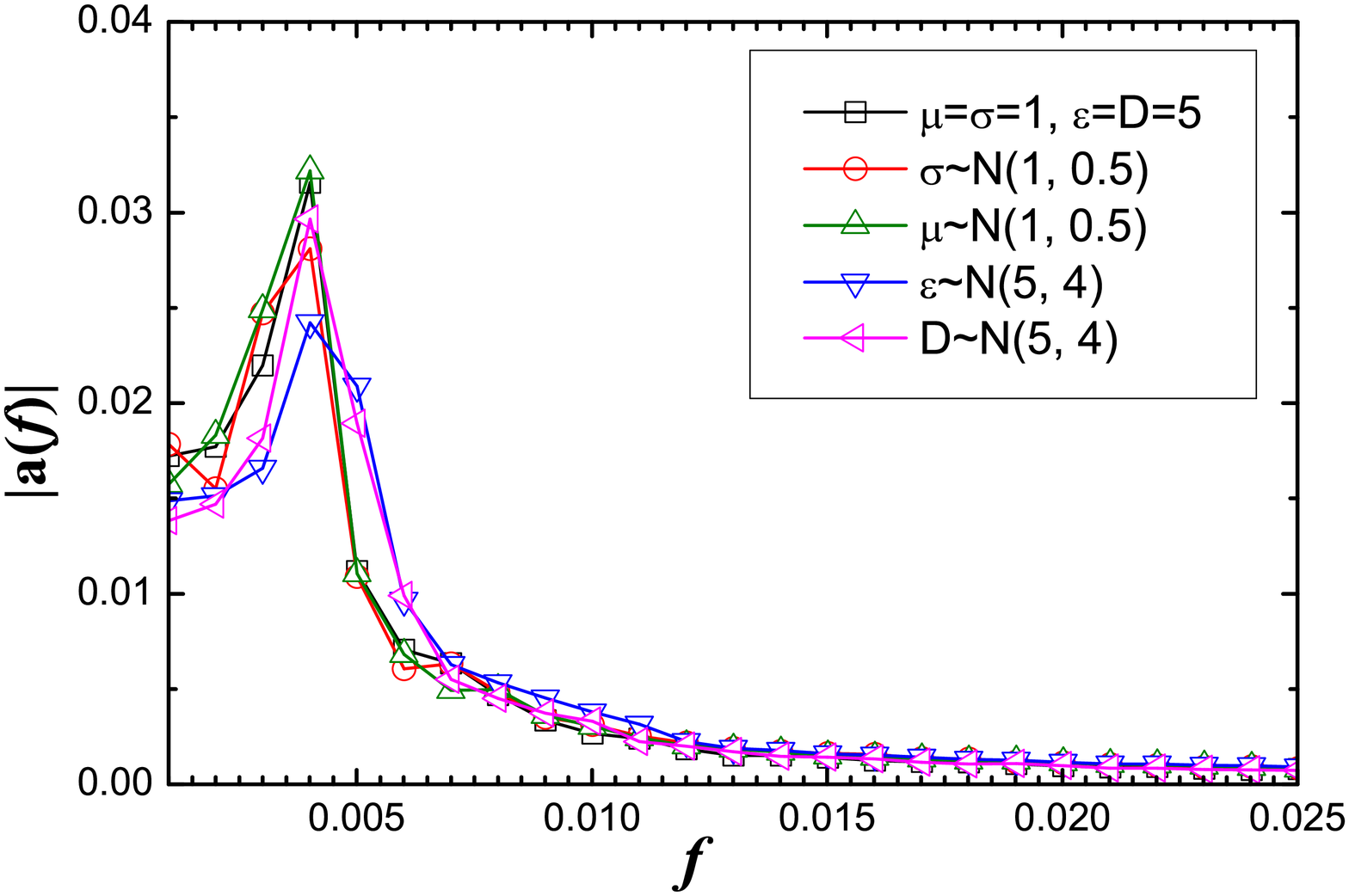}} \\
\subfloat[Correlation function $C_{AA}(x)$.]{\label{fig2:Caa}
\includegraphics[width=0.43\textwidth]{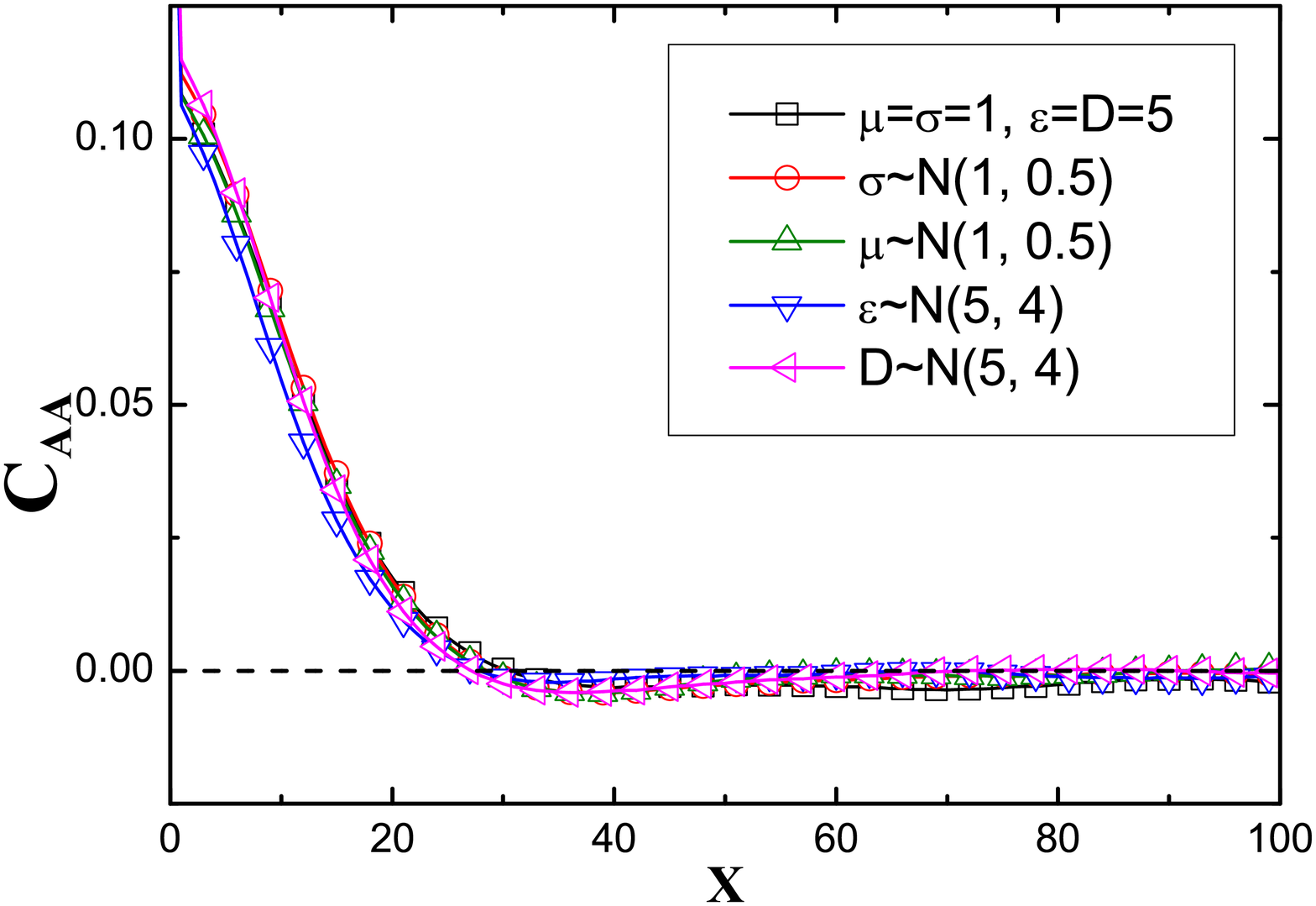}} 
\subfloat[Cross-correlation $C_{AB}(x)$.]{\label{fig2:Cab}
\includegraphics[width=0.43\textwidth]{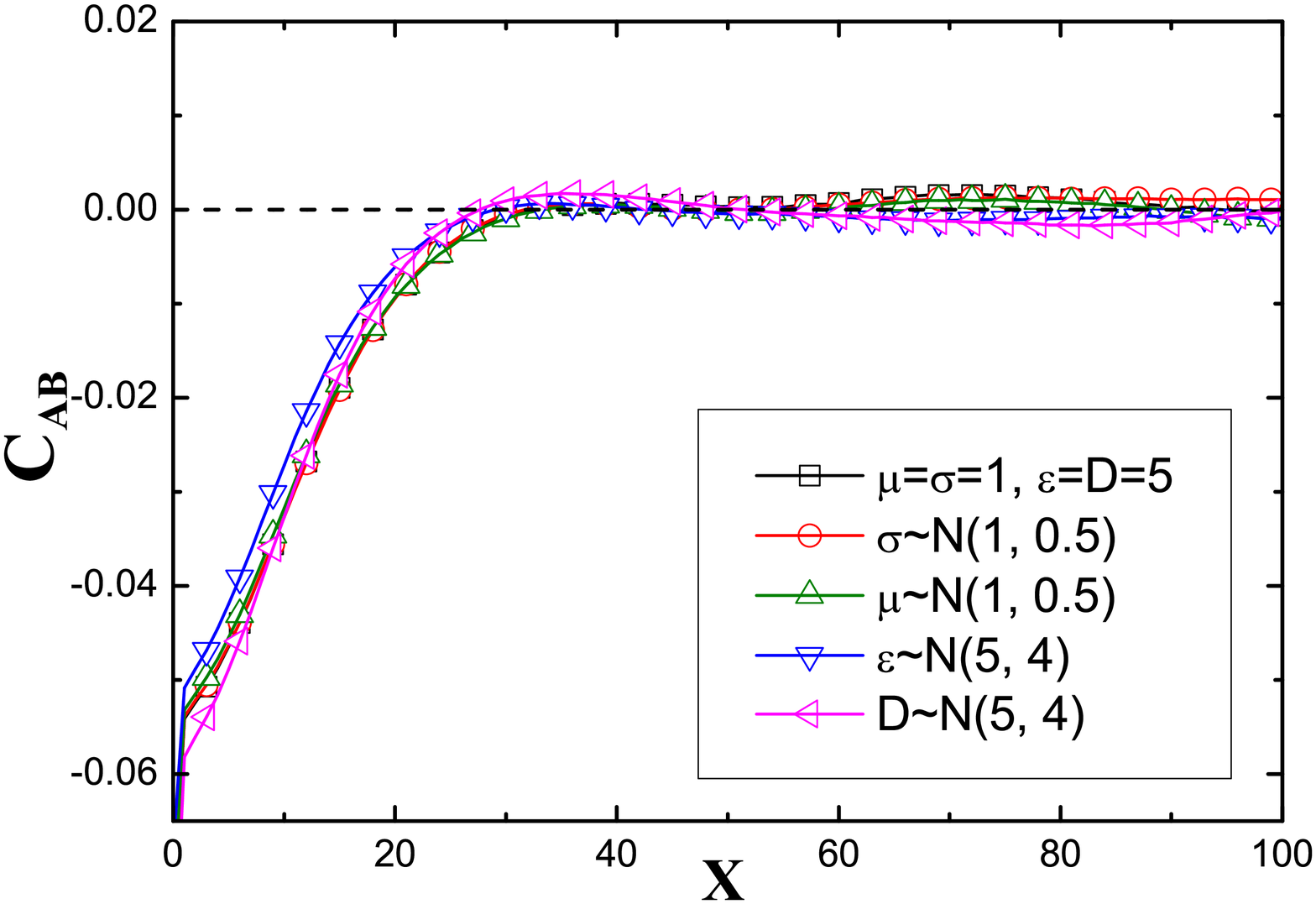}}
\end{center}
\caption{{\it (Color online.)} Quantitative observables for a stochastic
  May--Leonard system with $N = 256 \times 256$ sites, starting with equal 
  initial densities $a(0) = b(0) = c(0) = 1/3$ and in the presence of spatial 
  disorder, averaged over 50 simulation runs.
  If unspecified, the default rate values $\mu = \sigma = 1$, 
  $\epsilon =  D = 5$ were used.
  The correlation functions in (c) and (d) were measured at $t = 1000$ MCS.}
\end{figure*}
In order to quantitatively characterize the properties of the system, the 
influence of quenched spatial disorder, and the effect of pure particle pair 
exchange processes on the evolution of system, we next depict the temporal 
evolution of population density $a(t)= \langle n_A(j,t) \rangle$, the 
associated Fourier transforms $a(f)$, and the spatial auto- and 
cross-correlation functions $C_{AA}$ and $C_{AB}$, respectively, in the 
quasi-stationary state (here, at time 1000 MCS) in Fig.~2.  
Since the reaction processes are symmetric with respect to the three species,
the quantities associated with species $A$ suffice to extract the relevant 
information for all three species. 
As shown in Fig.~\ref{fig2:density}, the population density decreases swiftly 
at the beginning of the simulation runs: as the particles are initially 
randomly distributed and fill the entire lattice, predation reactions 
(\ref{react1}) dominate and deplete the particle density at the beginning. 
However, with the emergence of spiral patterns, these reactions can only take 
place along the domain boundaries of distinct species.
The system then evolves towards the (quasi-)steady state with population 
densities $\approx 0.26$ for the set of rates chosen here, consistent with the 
mean-field prediction $a^* = \frac{\mu}{3 \mu + \sigma} = 0.25$. 

In Fig.~\ref{fig2:FFT}, we depict the amplitude of the Fourier signal of the
population density.
The peak in $|a(f)|$ yields a characteristic oscillation frequency 
$\approx 0.004$.
This is almost a factor ten smaller than the prediction from the mean-field 
approximation, $f = \omega / 2 \pi = \sqrt{3} / 16 \pi \approx 0.034$, 
indicating a strong downward renormalization as consequence of spatial
fluctuations and correlations, similar to the situation in the stochastic
Lotka--Volterra model \cite{Ivan,Mark} but in stark constrast with the 
conserved spatially extended RPS model \cite{Qian}.
The finite width of the frequency peak in the Fourier plot indicates that the 
population oscillations will decay and ultimately cease after a finite 
relaxation time, consistent with the damped density fluctuations visible in
Fig.~\ref{fig2:density}.
Therefore, in the coexistence phase, the system's dynamics is consistent with 
the mean-field description, a feature that is however caused by the important
influence of the particles' spatial mobility: 
With relatively low but still effective (average) mobility rates (on average 
$\epsilon = D = 5$), the dynamics of the system is dominated by local 
interactions (reproduction and predation) along the boundaries of local spiral 
clusters.
Thereby, the coexistence state is maintained for a very long time and 
simultaneously effective mobility mixes the system well, resulting in a 
remarkably faithful description of the system through the mean-field 
approximation.
Furthermore, the (quasi-)stationary auto- and cross-correlation functions in 
Figs.~\ref{fig2:Caa} and \ref{fig2:Cab} decrease from their extremal values at 
vanishing distance to zero within about $\ell = 20$ lattice sites. 

More importantly, Fig.~2 shows the influence of spatial disorder on the 
physical quantities in the May--Leonard system.
We find that quenched randomness in the rates does not noticeably affect the 
temporal evolution of the population densities, the associated Fourier 
transform signals, or the decay lengths of the auto- and cross-correlation 
functions, irrespective of which rate is taken as random variable.
This result demonstrates that our previous observation in the four-state RPS 
model with conservation law, where we found spatial disorder to have only minor
effects, is also valid for the three-species May--Leonard system.
Therefore, we conclude that predator--prey systems with cyclic competition 
appear to be generically robust against random spatial variations in the 
predation, proliferation, or mobility rates.

\begin{figure*}[!t]
\label{Fig3}
\begin{center}
\subfloat[Temporal evolution $a(t)$.]{\label{fig3:density}
\includegraphics[width=0.43\textwidth]{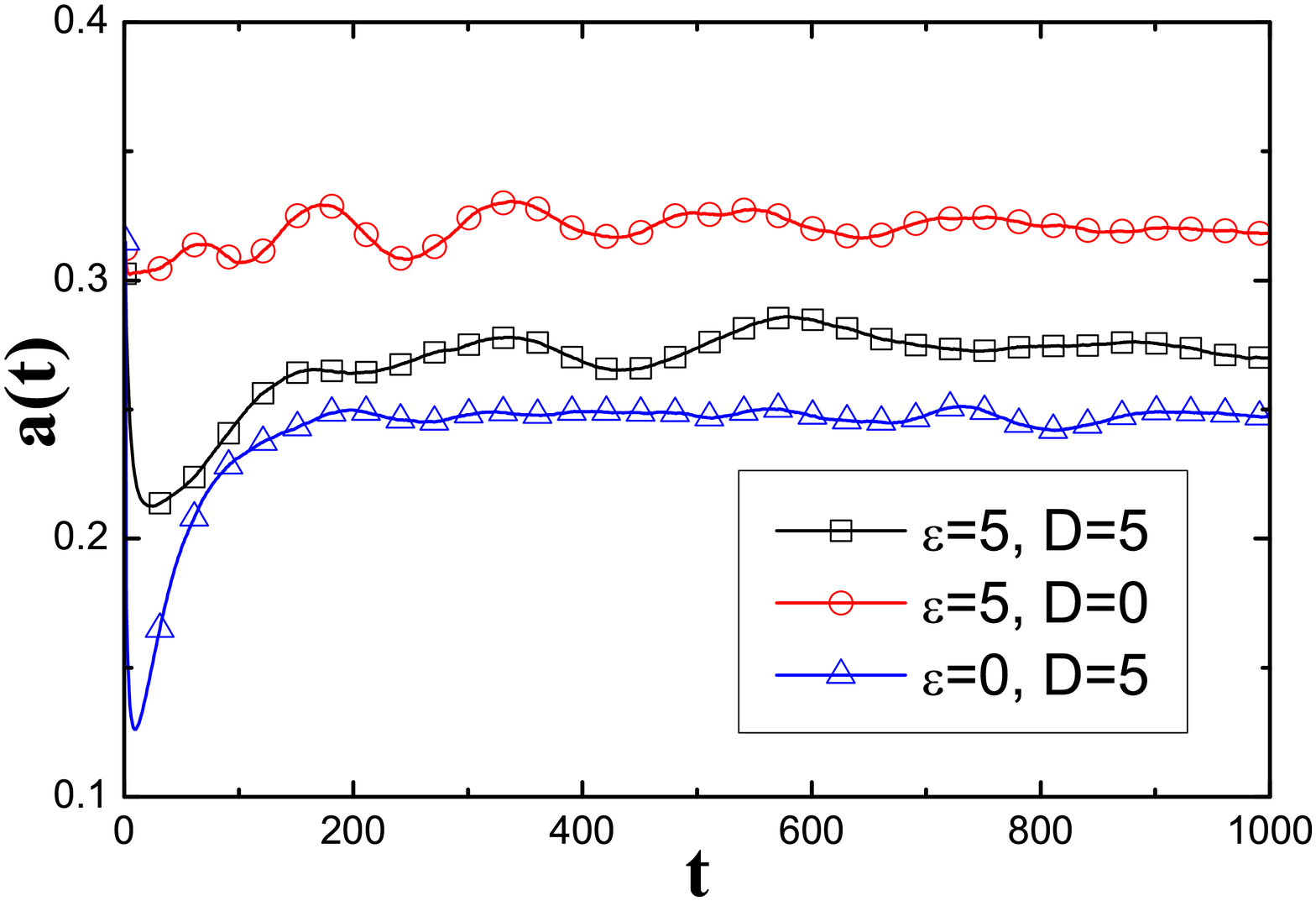}} 
\subfloat[Fourier transform $|a(f)|$.]{\label{fig3:FFT}
\includegraphics[width=0.43\textwidth]{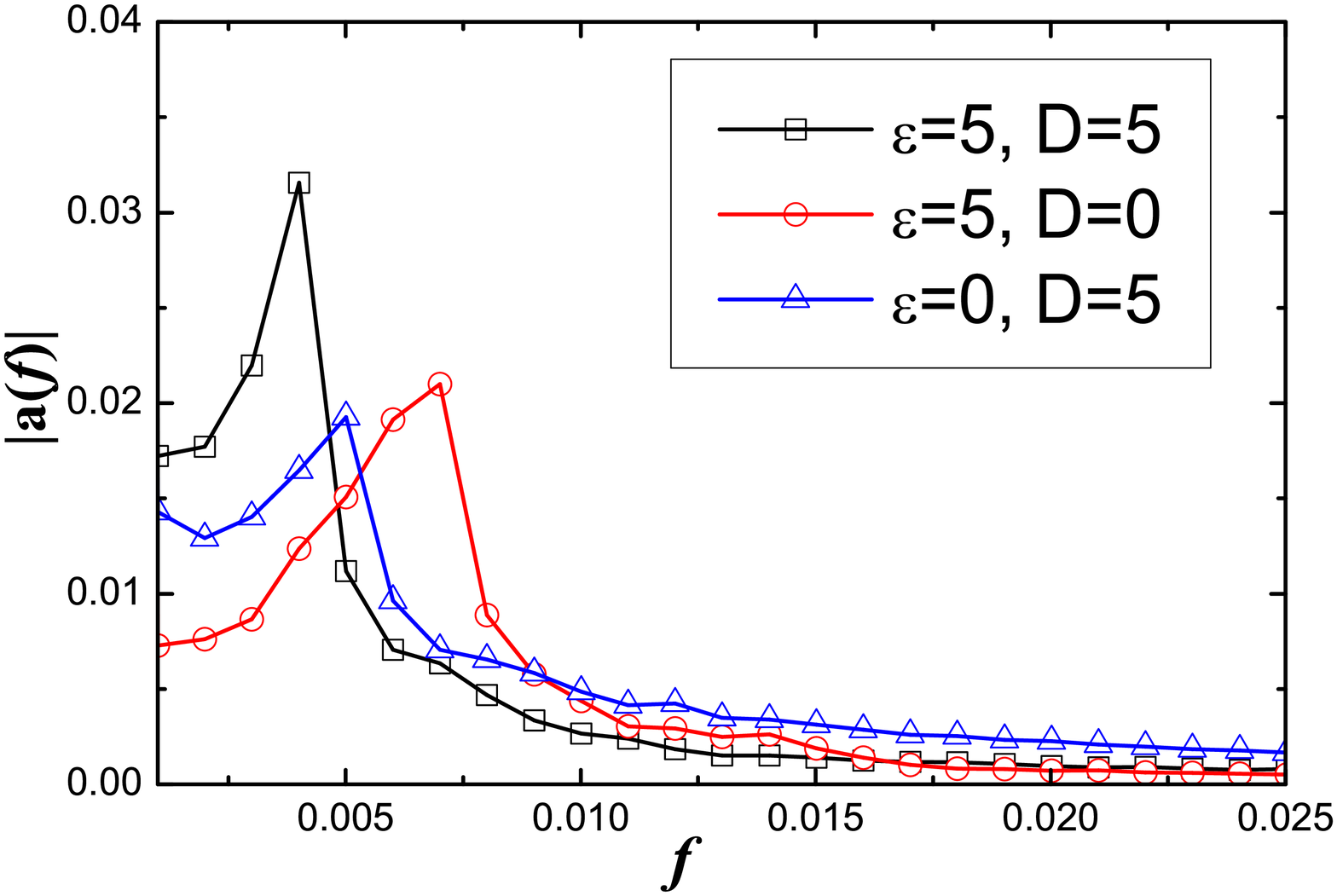}} \\
\subfloat[Correlation function $C_{AA}(x)$.]{\label{fig3:Caa}
\includegraphics[width=0.43\textwidth]{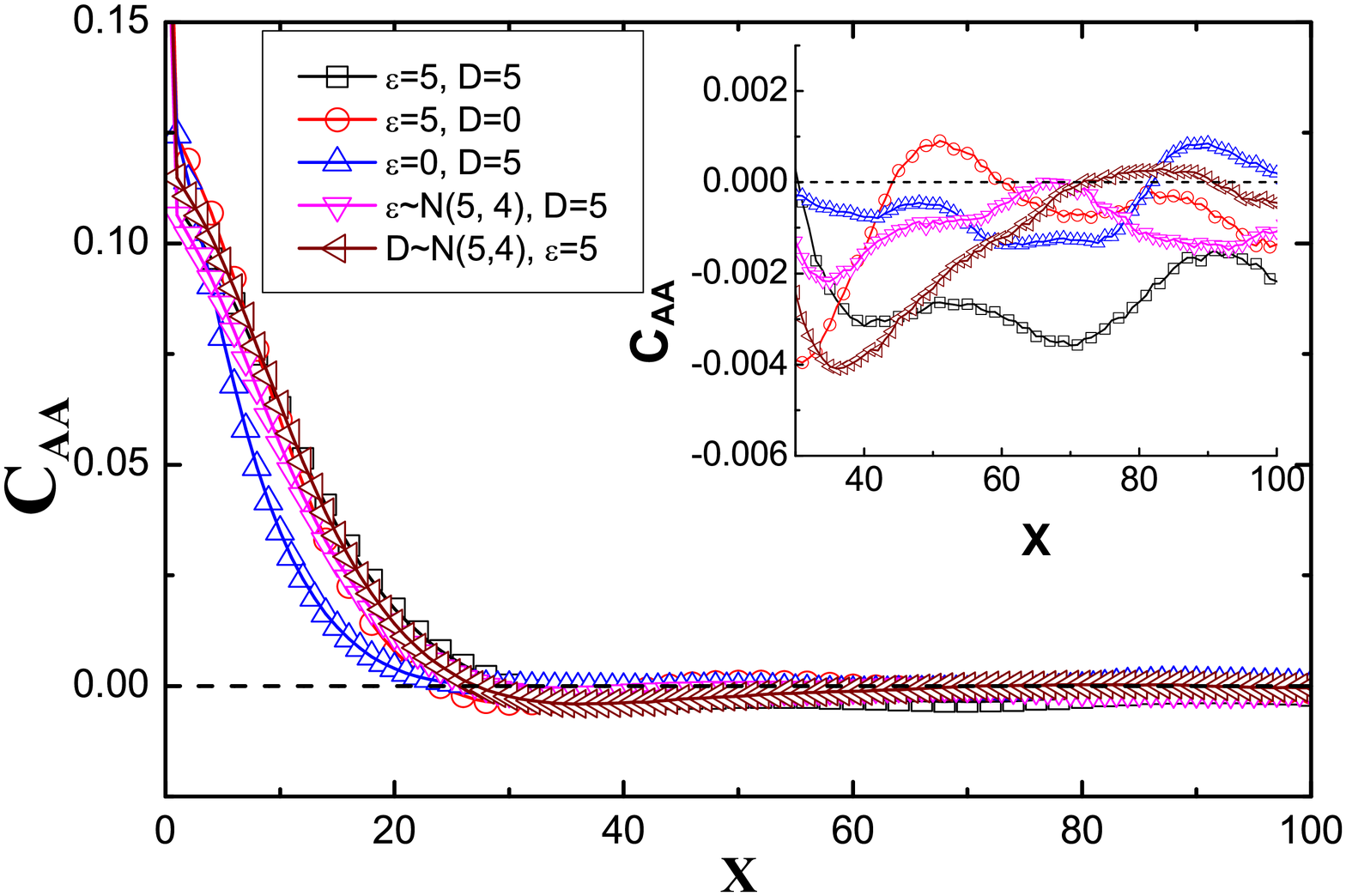}} 
\subfloat[Cross-correlation $C_{AB}(x)$.]{\label{fig3:Cab}
\includegraphics[width=0.43\textwidth]{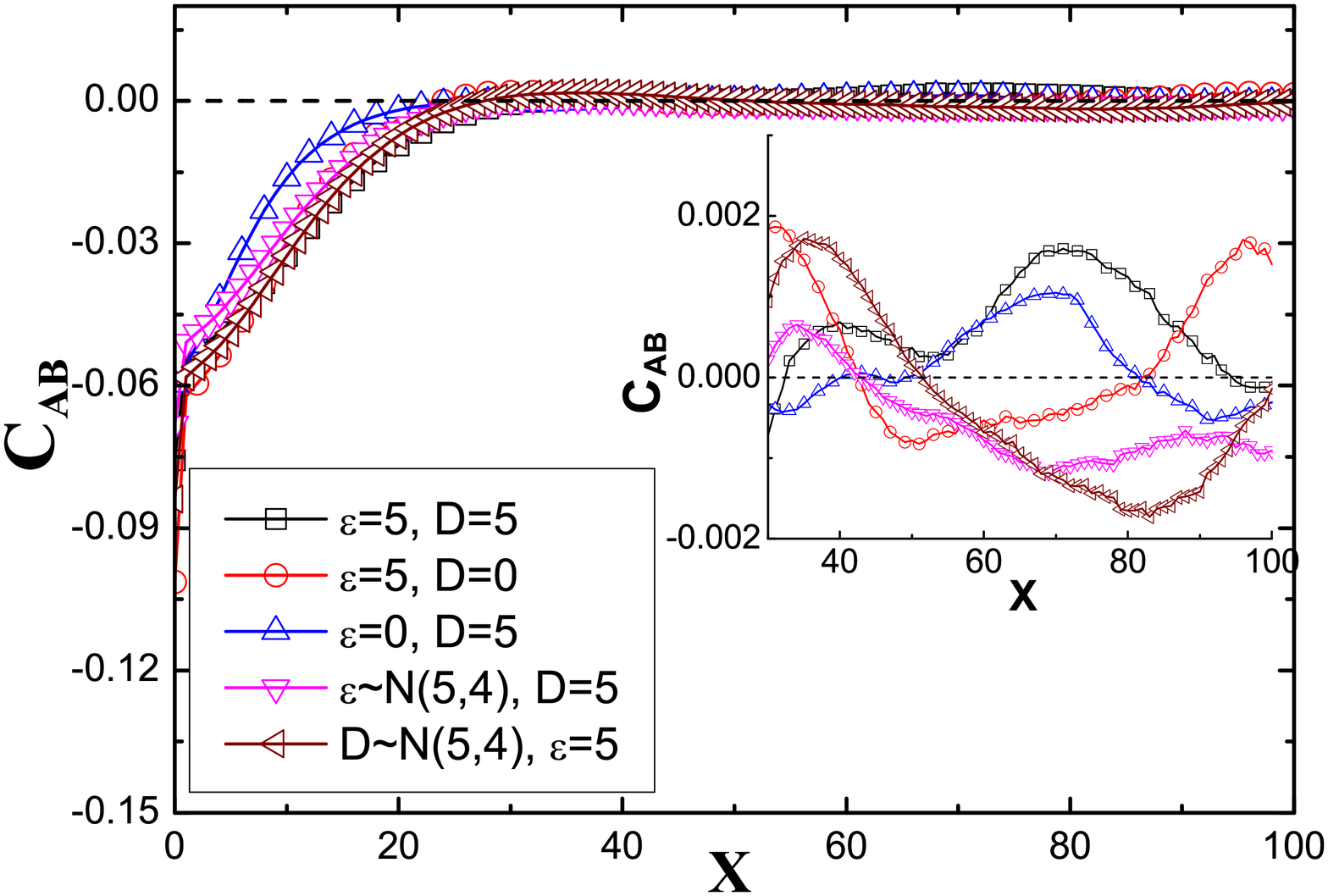}}
\end{center}
\caption{{\it (Color online.)} Quantitative observables for a stochastic
  May--Leonard system with $N = 256 \times 256$ sites, reaction rates 
  $\mu = \sigma = 1$, and starting with equal initial densities 
  $a(0) = b(0) = c(0) = 1/3$, with different combinations of nearest-neighbor 
  particle exchange and hopping processes, averaged over 50 simulation runs. 
  The correlation functions in (c) and (d) were measured at $t = 1000$ MCS.}
\end{figure*}
As demonstrated in Fig.~3, the formation of cluster patterns (see the third row
of Fig.~1) ultimately renders the various observables qualitatively similar to 
those systems in which particles can move only via exchange processes.
However, in Fig.~3 we also observe that the spatially averaged 
(quasi-)stationary density $a^*$ and the peak in the Fourier transform $|a(f)|$
vary according to the degree to which nearest-neighbor hopping is included in 
the model. 
In particular, when particle dispersal happens solely through hopping processes
(\ref{hopp}), the asymptotic species densities are relatively low, see 
Fig.~\ref{fig3:density} with $\epsilon = 0$, $D=5$; compare with the two plots 
for other rate choices, and the corresponding snapshots \ref{fig1:snpsh2}, 
\ref{fig1:snpsh7}, and \ref{fig1:snpsh8}.
We attribute this lower overall species fitness to the previously noted 
emergence of sizeable clusters of just empty lattice sites in the absence 
of pair exchange processes (\ref{exch}), visible as small black patches in 
Figs.~\ref{fig1:snpsh7} and \ref{fig1:snpsh9}.
In fact, these voids effectively buffer the three species against the predation
reactions (\ref{react1}), but also diminish the total area that can be 
saturated by either population, which results in an overall density reduction 
in the (quasi-)steady state.
In contrast, when particle pair exchange processes are included, the influence 
of empty-site clusters is diminished and consequently the (quasi-)stationary 
population densities enhanced (see Fig.~\ref{fig3:density} for 
$\epsilon \neq 0$). 
Furthermore, we observe that the characteristic density Fourier peak 
frequencies in systems where nearest-neighbor hopping is allowed are 
renormalized to even lower values than in runs with just pair exchange 
processes, which tend to better mix the system, see Fig.~\ref{fig3:FFT} 
(compare the characteristic frequency for the runs with $D = 5$ to those with 
$D = 0$). 
In Figs.~\ref{fig3:Caa} and \ref{fig3:Cab}, we observe the presence of 
low-amplitude population (damped) oscillations during the decay of the 
correlation function, which originate from the spiral structures displayed by 
the system in the coexistence state (under ``efficient'' mobility rates) 
\cite{Reichenbach4}. 
In the insets of Figs.~\ref{fig3:Caa} and \ref{fig3:Cab}, we notice that the 
correlation functions of the model variants with only hopping processes 
($\epsilon = 0$) decay to zero in a less oscillatory manner than the correlation 
functions for the variants that include particle exchange ($\epsilon \neq 0$).
This is a consequence of the observed absence of well-defined spiral structures
(see Figs.~\ref{fig1:snpsh7} and \ref{fig1:snpsh9}) when the pair-exchange rate
is too low to efficiently stir the system.

In summary, the pair-exchange processes suppress the presence of empty-site 
clusters, and despite rendering the spiral structures more diffuse (i.e., more 
``entangled''), they have an overall stabilizing effect on emerging spatial 
patterns, and consequently promote the fitness and coexistence of all three 
subpopulations $A$, $B$ and $C$.

\subsection{Mean extinction times and their distribution}

As we have discussed above, the system is efficiently stirred when the pair 
exchange rate $\epsilon$ is high enough (independent of the actual value of 
$D$). 
In this case, the system is characterized by a long-lived coexistence state. 
However, when the pair exchange rate is below some critical value, it has been 
shown that the system settles in an absorbing state after an observable amount 
of time \cite{Reichenbach1}.  
We here revisit and extend the analysis of such a scenario that holds true for 
any values of $\epsilon$ or $D$ by computing the mean extinction time (MET) and
the distribution of extinction times. 
Here, for the sake of simplicity (and without loss of generality) we assume 
that $\epsilon = D$ and the mobility rate therefore is $M = 2\epsilon / N$. 
In this setting, our May--Leonard model coincides with the variant considered 
in Refs.~\cite{Reichenbach1,Reichenbach2,Reichenbach4}, where it was shown that
the critical mobility threshold is $M_c \approx 4.5 \times 10^{-4}$ (when 
$\mu=\sigma=1$).
Indeed, when the effective mobility rate $M$ approaches $M_c$ from below, there
is a cross-over from a coexistence (quasi-)steady state to an absorbing state.

\begin{figure*}
\label{Fig4}
\begin{center}
\subfloat[$\bar{T}_{\rm ex}$ vs. $N$.]{\label{fig4:MET1}
\includegraphics[width=0.33\textwidth]{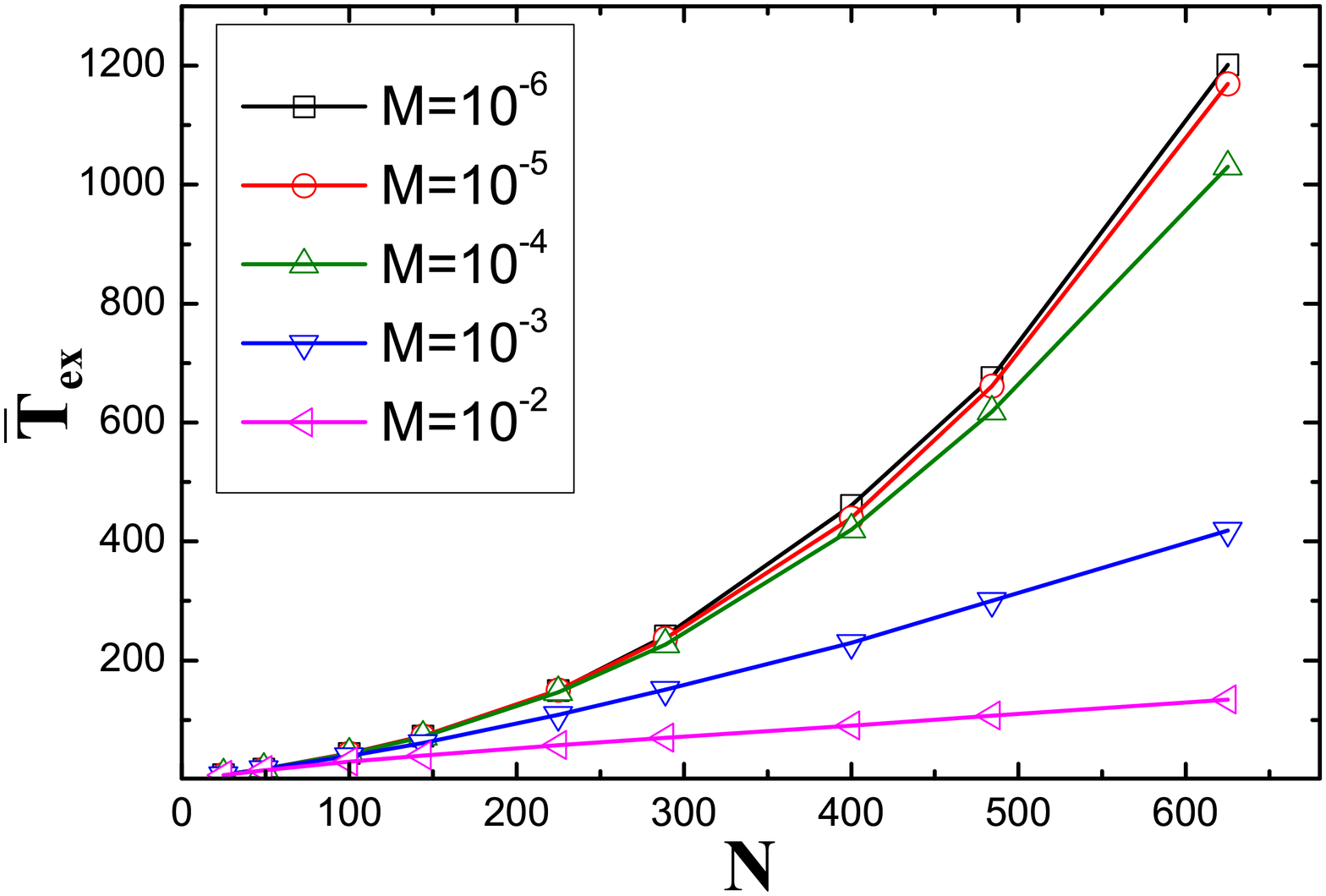}}
\subfloat[$\log_{10}(\bar{T}_{\rm ex}\times N)$ vs. $N$.]{\label{fig4:MET2}
\includegraphics[width=0.33\textwidth]{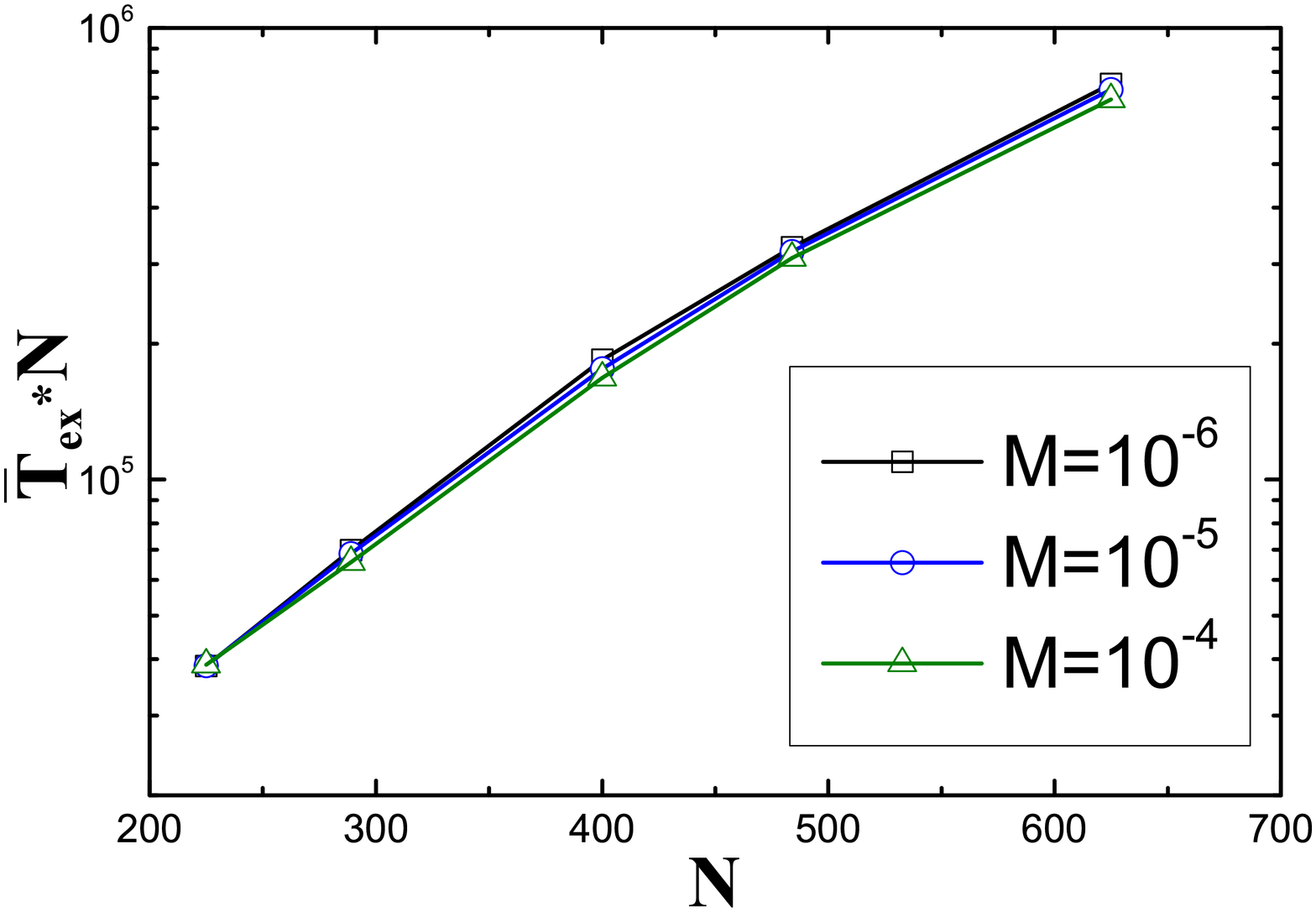}}
\subfloat[$\bar{T}_{\rm ex}$ vs. $N$.]{\label{fig4:MET3}
\includegraphics[width=0.33\textwidth]{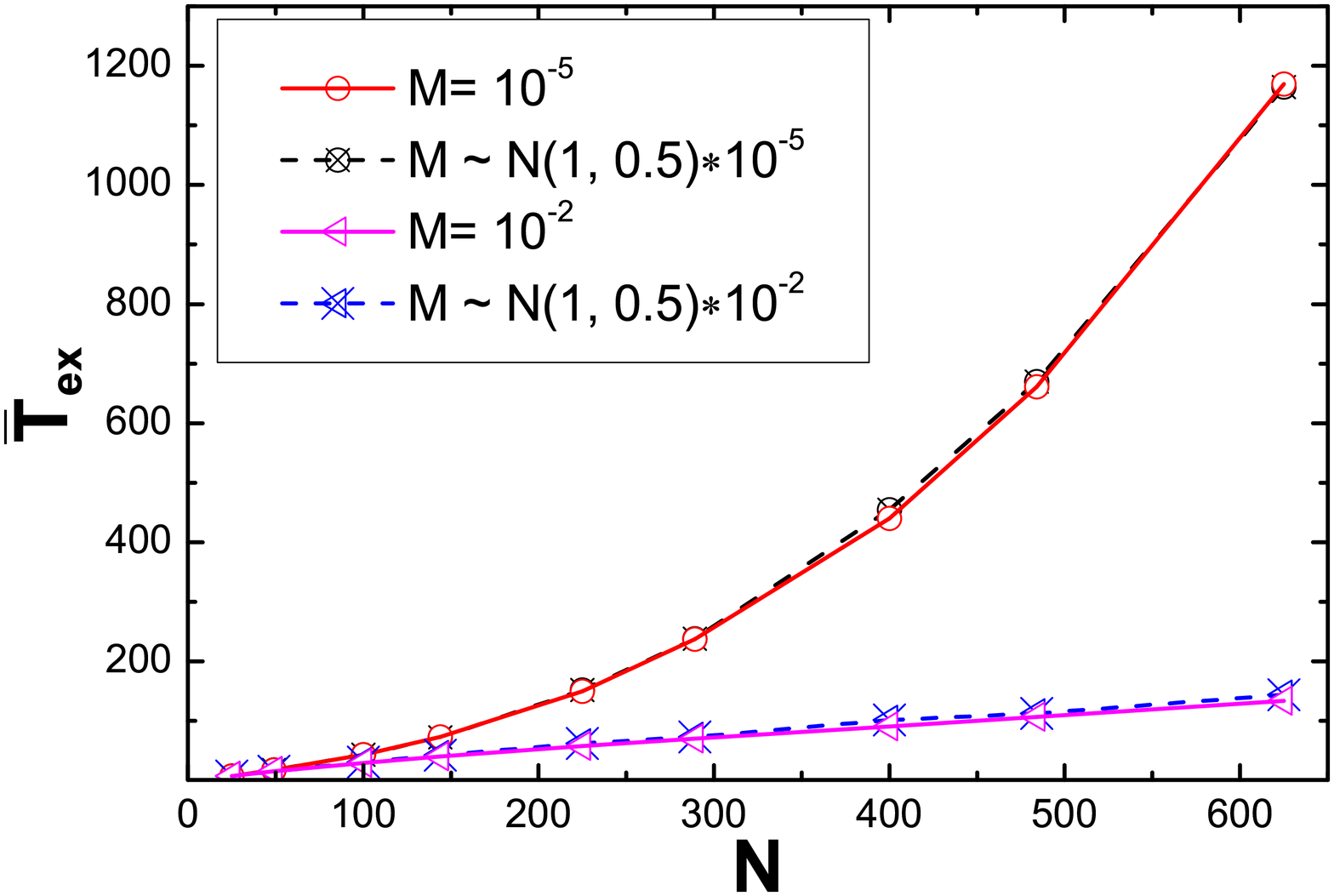}}\\
\subfloat[$\epsilon = D = 0.1$; inset: $\epsilon=D\sim N(0.1, 0.05).$]
  {\label{fig4:dist1}
\includegraphics[width=0.43\textwidth]{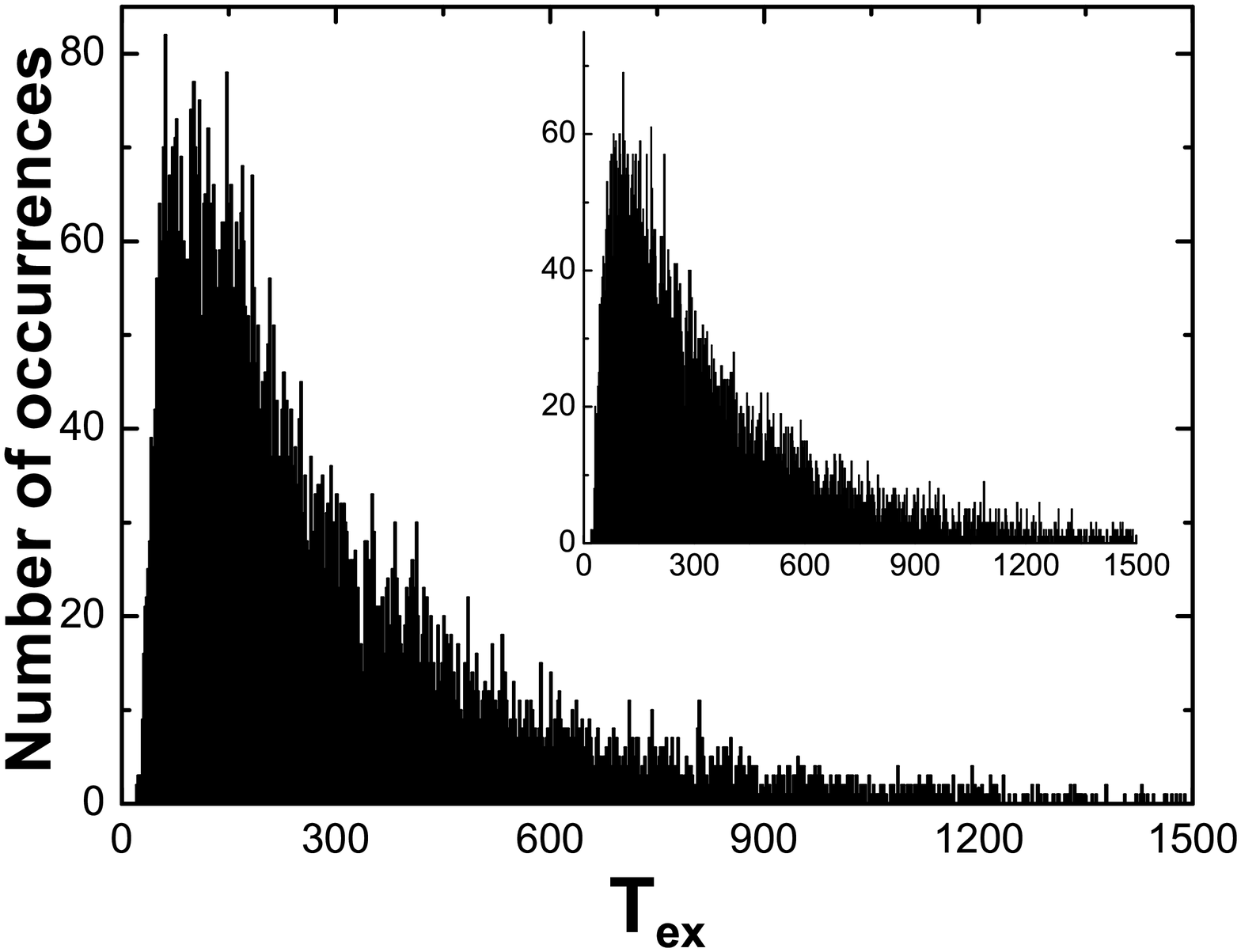}}
\subfloat[$\epsilon = D = 25$; inset: $\epsilon=D\sim N(25, 10)$.
  The dashed (red/gray) curve depicts a Gaussian fit.]{\label{fig4:dist2}
\includegraphics[width=0.43\textwidth]{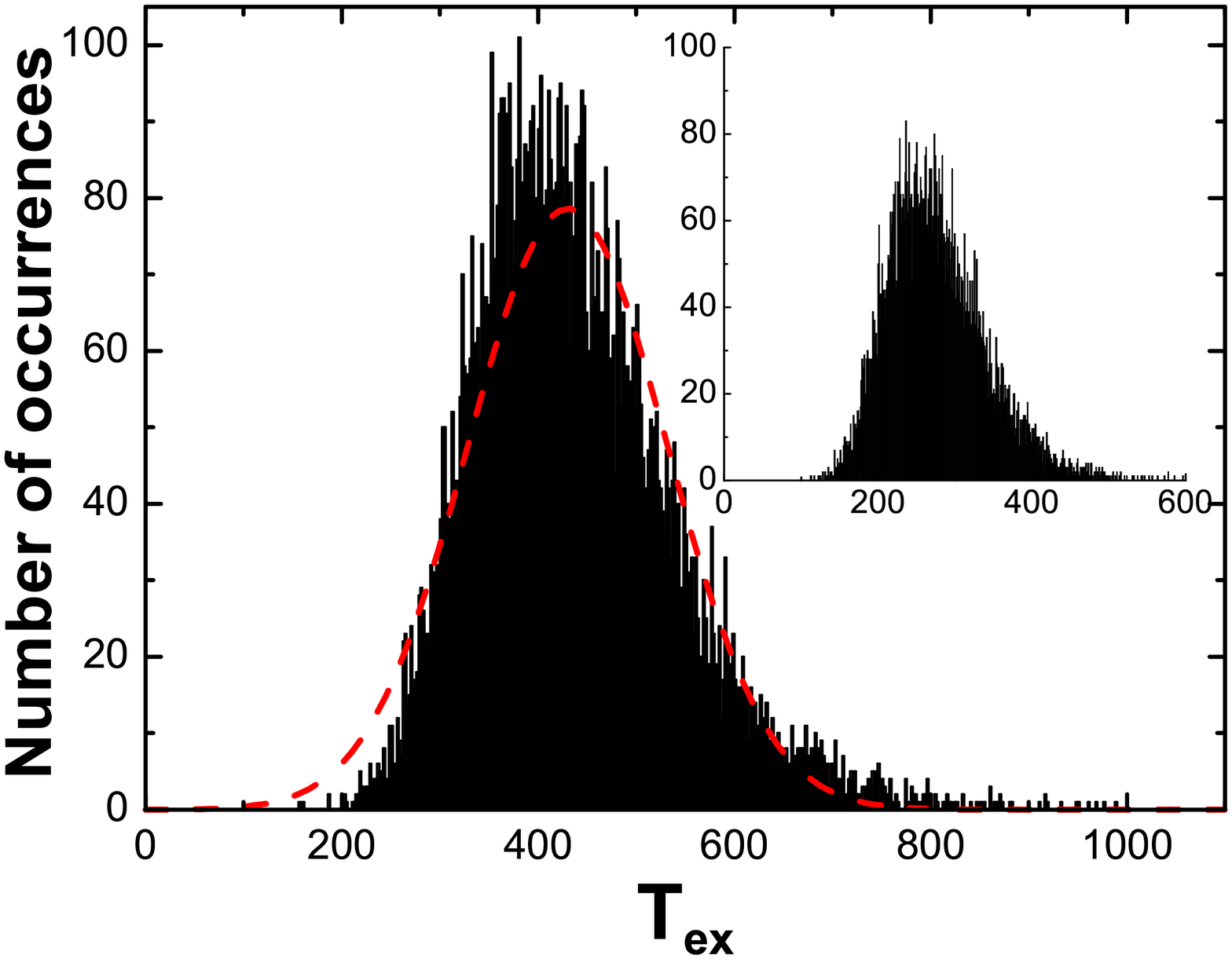}}
\end{center}
\caption{{\it (Color online.)} (a), (b), (c) Mean extinction time (MET) 
  $\bar{T}_{\rm ex}$ as function of lattice size $N$, for different values of 
  the effective mobility $M = 2 \epsilon / N$ (here, $\epsilon = D$), obtained 
  from averages over 10000 Monte Carlo runs, starting with equal initial 
  densities $a(0) = b(0) = c(0) = 1/3$, and reaction rates $\mu = \sigma = 1$. 
  The lattice sizes are $N = 5 \times 5$, $7 \times 7$, $10 \times 10$, 
  $12 \times 12$, $15 \times 15$, $17 \times 17$, $20 \times 20$, 
  $22 \times 22$, $25 \times 25$ sites. \ 
  (d) and (e): Histogram of extinction times estimated from 10000 Monte Carlo 
  runs, based on a system with lattice size $N = 20 \times 20$. 
  The insets correspond to the histograms obtained for random mobility rates:
  $\epsilon=D\sim N(0.1, 0.05)$ in (d) and $\epsilon=D\sim N(25, 10)$ in (e).}
\end{figure*}

The MET has been computed as the time when the first of the three species dies 
out.
Figure~4 illustrates how this MET exhibits markedly different behavior when the
effective mobility rate $M=2\epsilon/N$ is swept through $M_c$.
As shown in Fig.~\ref{fig4:MET2}, when the mobility is relatively weak 
($M = 10^{-6}$), the MET increases with the system size approximately according
to the (zero-dimensional) functional form 
$\bar{T}_{\rm ex}(N) \sim e^{{\rm c} N}/N$ (where c is a constant) 
\cite{mauro}, especially for comparatively large values 
$N \sim 200 \ldots 600$.  
(For smaller systems with $N < 200$, the data do not fit this functional
dependence very well, but may also not be as statistically reliable.)
Yet the curvature of the graphs in Fig.~\ref{fig4:MET1} decreases upon raising 
the effective mobility, and the functional dependence on system size becomes 
replaced with a linear form $\bar{T}_{\rm ex}(N) \sim N$ for $M > M_c$.
That is, when the mobility rate is low, the system is dominated by local 
interactions and species extinction is a rare event driven by a large 
fluctuation after an enormous amount of time. 
In this case, the coexistence of the three species corresponds to a metastable 
state. 
Interestingly, Fig.~\ref{fig4:MET3} shows that spatial disorder in the mobility
rate $M$ does not qualitatively affect the behavior of the MET: this very same
scenario applies even when $M$ is randomly distributed.
This observation further supports the conclusion that spatial variability in 
the mobility rates has little effect on the dynamical evolution of the system.

When the mobility rate increases and exceeds the threshold, the system is 
regularly driven towards extinction and biodiversity is lost, just as predicted
by the zero-dimensional formulation of the model.
Furthermore, the histograms of extinction times plotted in 
Figs.~\ref{fig4:dist1} and \ref{fig4:dist2} show that the extinction time
distributions (obtained for small systems with $N = 20 \times 20$ sites)
correspondingly evolve gradually from an approximately exponential (or Poisson)
shape, albeit with fat tails, towards a (roughly) Gaussian distribution 
centered at $\bar{T}_{\rm ex}$. 
In addition, the distributions remain unchanged even if spatial disorder is 
incorporated in the model (see the insets of Figs.~\ref{fig4:dist1} and 
\ref{fig4:dist2}).
The approximate  quasi-exponential (or quasi-Poisson) distribution of 
Fig.~\ref{fig4:dist1} is a characteristic feature of systems where extreme 
events occur only after a very long time, and which are hence driven by large 
rare fluctuations \cite{Touchette}.
Here, the rare extreme event is extinction of species that were previously 
coexisting in a metastable state for a long time period. 
On the other hand, the approximately Gaussian distribution of 
Fig.~\ref{fig4:dist2} is typical of systems where random fluctuations are of 
weak intensity \cite{Gardiner}.
In our competing three-species system, it is associated with the absorbing 
state that corresponds to species extinction happening within an ``observable''
time range, as in the zero-dimensional counterpart of the model
\cite{Reichenbach1,Reichenbach2,Reichenbach4}.
These numerical observations support the method suggested in 
Ref.~\cite{Reichenbach1} for identifying the nature of (quasi-)steady states in
the simulation of RPS models: 
A reasonable criterion is that the system remains at the steady state if three 
species still coexist after simulation time $t \sim N$, otherwise, the system 
evolves to an absorbing state.

\section{Conclusion}
\label{conclu}

In this paper, we have demonstrated that quenched spatial disorder in either 
the reaction or the mobility rates does not significantly affect the temporal 
evolution, Fourier signals, spatial correlation functions, or mean 
extinction times in stochastic spatial May--Leonard models (i.e., four-state 
RPS models without total particle number conservation) in two dimensions.
In combination with our previous results for conserved three-species RPS 
systems \cite{Qian}, we conclude that such cyclic predator-prey systems appear
to be generically robust against spatial variability of the rates.
Here, the randomized reaction rates remain attached to the lattice sites, 
mimicking environmental variations that do not change over time. 
As there exist a number of systems, e.g. in ecology \cite{Sinervo} and 
microbiology \cite{Kerr}, where the competition among species is cyclic, an 
important implication of our findings is that the environmental variability of 
the parameters can essentially be neglected in the mathematical description of 
those systems.
In addition, through removing the hopping process by letting $D=0$, we observe 
that particle pair exchange processes promote the formation of sharp spiral 
patterns.
In our spatial stochastic system, we measure the population oscillation 
frequency to be much lower than predicted by the mean-field rate equations,
similar to the situation in two-species Lotka--Volterra models 
\cite{Ivan,Mark}, but in stark contrast with our numerical results for 
conserved RPS model variants \cite{Qian}.
This downward frequency renormalization is enhanced by the presence of 
nearest-neighbor hopping processes.
Moreover, we find a remarkable gradual transformation in the dependence of the
mean extinction time on system size, and the shape of the associated extinction
time distribution, when the effective mobility rate crosses the critical 
threshold separating the coexistence from the absorbing state: 
When the mobility rate is low, the distribution of extinction times is 
approximately exponential, and species coexistence corresponds to a 
long-lived metastable state. 
In this case extinction is driven by large, rare fluctuations and the mean 
extinction time essentially grows exponentially with the population size.
Above the critical mobility threshold, the extinction times are approximately 
distributed according to a Gaussian.
In this situation, the noise is of weak intensity and the mean extinction time
grows linearly with the population size. 
Interestingly, we find that these results remain valid for both non-random as 
well as for randomly distributed mobility rates.

\noindent 
This work is in part supported by Virginia Tech's Institute for Critical 
Technology and Applied Science (ICTAS) through a Doctoral Scholarship. 
We gratefully acknowledge inspiring discussions with George Daquila, Uli 
Dobramysl, Michel Pleimling, Matt Raum, and Royce Zia.

\end{document}